\documentclass[11pt]{article}
\pdfoutput=1
\usepackage[margin=1in]{geometry}
\usepackage[skip=0pt]{parskip}  
\usepackage{graphicx}
\usepackage{amssymb}
\usepackage{dsfont}
\usepackage{mathtools}
\usepackage{physics}
\usepackage[english]{babel}
\usepackage{qcircuit}
\usepackage[font=small]{caption}
\usepackage{nicefrac}
\usepackage{listings}
\usepackage{color}
\usepackage{hyperref}
\usepackage{tikz}
\usetikzlibrary{angles,quotes}
\usepackage{tkz-euclide}
\usepackage{fancyhdr}
\usepackage{booktabs}
\usepackage{arydshln}

\definecolor{dkgreen}{rgb}{0,0.6,0}
\definecolor{gray}{rgb}{0.5,0.5,0.5}
\definecolor{mauve}{rgb}{0.58,0,0.82}
\definecolor{dkblue}{rgb}{0,0.3,0.6}
\definecolor{dkred}{rgb}{0.6,0,0}
\definecolor{purple}{rgb}{0.48,0.13,0.44}
\definecolor{dkgray}{rgb}{.25,.25,.25}

\newcommand{\RR}{\mathds{R}}   
\newcommand{\CC}{\mathds{C}}   
\newcommand{\NN}{\mathds{N}}
\newcommand{\PP}{\mathds{P}}   
\newcommand{\EE}{\mathds{E}}   
\newcommand{\VV}{\textrm{Var}}   
\newcommand{\indic}{\mathds{1}}   

\title{\textbf{Potential Applications of Quantum Computing for the Insurance Industry}}

\author{Michael Adam\thanks{\texttt{michael.adam@axa.de} } \hspace{3pt}\\
	\small{\textit{AXA Konzern AG\thanks{in collaboration with Fraunhofer ITWM}}}
}

\date{\small{October 10, 2022}}                                          

\begin{document}
\maketitle

\begin{abstract}
This paper is the documentation of a pre-study performed by AXA Konzern AG in collaboration with Fraunhofer ITWM to assess the relevance of quantum computing for the insurance industry. Beside a general overview of the status quo of quantum computing technologies, we investigate its applicability for the valuation of insurance contracts as a concrete use case. This valuation is a computationally intensive problem because the lack of closed pricing formulas requires the use of Monte Carlo methods. Therefore current technical capabilities force insurers to apply approximation methods for many subsequent tasks like economic capital calculation or optimization of strategic asset allocations. The business-criticality of these tasks combined with the existence of a quantum algorithm called Amplitude Estimation which promises a quadratic speed-up of Monte Carlo simulation makes this use case obvious. We provide a detailed explanation of Amplitude Estimation and present two quantum circuits which describe insurance-related payoff features in a quantum circuit model. An exemplary circuit that encodes dynamic lapse is evaluated both on a simulator and on real quantum hardware.
\end{abstract}

\newpage
\tableofcontents
\newpage

\section{Introduction}
Quantum computers work fundamentally different than classical computers. By leveraging the laws of quantum mechanics, they promise significant speed-ups for certain problems. The areas of potential applications are versatile and include  i.a. cryptography, machine learning and simulation. This opens up several possibilities for the insurance industry to benefit from the new technology.\\ 

Broadly speaking, quantum computers can help solving computationally intensive problems. When asking an actuary for a common task with high computational requirements, the valuation of insurance contracts is an obvious answer. The most expensive simulation concerns the liabilities of life and health companies with very long time horizons (40 years and beyond), complex contractual and legal frameworks, sophisticated customer behavior and interactions with assets (e.g. through profit participation). Due to the complexity of these "instruments", closed-form valuation is not available and hence Monte Carlo (MC) methods are widely used. But the fact is, that the valuation of liabilities is essential for many subsequent tasks like calculation of economic capital, stress test exercises or optimization of strategic asset allocations. Unfortunately, in many cases the current technical capabilities do not allow full nested simulations. If we for example consider economic capital, the full probability distribution of profits and losses is needed and hence MC simulations "within" MC simulations are necessary. Actually that often means that at least 10'000 so called outer scenarios and additionally not less than 1'000 inner scenarios for each outer are needed. Hence we easily reach millions of simulations with long time horizons.\\

In practice there are different approaches to reduce the total number of inner scenarios, in particular replicating portfolios \cite{boekel09} and least square Monte Carlo \cite{bauer12}. In both approaches a reduced number of simulations is processes and an "easy-to-valuate-proxy" is fit to the calculated information. The obvious drawbacks are (1) potential inaccuracies especially in extreme scenarios (which are often the most interesting ones), (2) additional work to generate the proxy and (3) additional effort to analyze the discrepancies. If quantum computing would speed-up the simulation of inner scenarios so that these proxies become obsolete, the impact on life and health insurers would be enormous: Besides reducing the effort due to the above-mentioned drawbacks, increasing calculation frequency could open the door to much more dynamic steering strategies. Import indicators like the Solvency 2 coverage ratio are for example currently calculated only a few times a year and consequently it is hard to use it for dynamic hedging or optimization purposes.\\

Recently published quantum algorithms promise a quadratic speed-up for Monte Carlo valuations of plain vanilla European options \cite{rebent18}, basket options \cite{stama20} as well as path-depended options like Asian \cite{rebent18} or Barrier options \cite{stama20}. I.e. the convergence rate increases from $1/\sqrt{M}$ to $1/M$ where $M$ is the number of samples. The underlying idea is \textit{Amplitude Estimation} \cite{brassard02}, where the target quantity is first encoded to a quantum circuit which is then manipulated so that a subsequent measurement delivers the desired result with high probability. Despite these promising results, the implementation of a realistic insurance model on a real hardware quantum computer cannot be expected in the upcoming years. Both hardware and software development are still in their infancy and many theoretical results can only be executed in their most basic version. On the other hand the development progresses and with the launch of IBM's Q System One in Germany another milestone was reached in June 2021.\\ 

In this documentation, we concentrate on quantum software development in a circuit model. In particular we implement a payoff representing a whole life insurance and a second one simulating stochastic customer behavior linked to interest rates. This is a contribution to the growing library of quantum circuits that can be used to model financial instruments or specifically insurance contracts. Before we start the description of the insurance-related circuits, we give a brief introduction to quantum computing in section \ref{sec:basics} followed by a general description of the building blocks needed for calculating expected values in section \ref{sec:calc_e}. Since amplitude estimation (AE) leads to the quadratic speed-up which is the main motivation for our work, we dedicate an own section to the derivation of AE (section \ref{sec:ae}). Then, after a mathematical formulation of the considered payoffs in section \ref{sec:payoff} we describe the new insurance-related quantum circuits in section \ref{sec:quantum_circuits}. In section \ref{sec:hardware_results} we show results from running parts of the dynamics lapse circuit on a simulator and on real quantum hardware. The paper finishes with a short overview of current quantum computing technology, both for hard- and software in section \ref{sec:current_tech}. 

\section*{Notation}
\begin{tabular}{l p{12cm}}
$\ket{\psi}, \ket{\varphi}, \ket{\rho}, \ket{\chi}, \ket{\upsilon}$ &
Qubit register in arbitrary superposition \\
$m, n, r, s$ &
Size of qubit registers. Corresponding number of basis states is denoted by capital letter, e.g. $M=2^m$ \\
$\ket{\psi}_m$ &
Index outside ket gives size of the qubit register in number of qubits. If size is $1$ or not relevant in the current context, the index is usually omitted.\\
$\ket{\psi_i}$ &
Index inside ket indicates $i^{th}$ qubit in register $\ket{\psi}_m$. If omitted entire register is meant. \\
$\ket{\psi_i}_{m_i}$ &
Index in- and outside ket indicates that $\ket{\psi}_m$ should be interpreted as register of registers, i.e. $\ket{\psi}_m=\ket{\psi_1}_{m_1}\otimes\dots\otimes \ket{\psi_n}_{m_n}$ with $\sum m_i=m$. Then $\ket{\psi_i}_{m_i}$ denotes  $i^{th}$ qubit register of $\ket{\psi}_m$. \\
$\ket{k}_m, \ket{l}_m$ &
Qubit in $k^{th}$ or $l^{th}$ basis state. $k,l\in\{0,\dots,2^{m-1}\}$ \\
$\alpha_k, \sqrt{a_k}$ &
Qubit Amplitudes \\
$i, t_i$ &
Usually used as time associated iterators \\
$\mathcal{A, B,}\dots$ &
Linear operators \\
$H,H^{\otimes m}$ &
Single and $m$-qubit Hadamard operator \\
$\vartheta,\theta,\phi,\gamma$ &
Angles \\
$H(\chi,\phi)$ &
Hyperplane spanned by $\ket{\chi}$ and its orthogonal space $\ket{\chi^\bot}$ with $\phi$ denoting the rotation angle in the complex dimension \\
$\lambda_k$ &
Eigenvalues \\
MC &
Monte Carlo \\
PE, AE, AA &
Phase Estimation, Amplitude Estimation, Amplitude Amplification \\
$QFT, QFT^{-1}$ &
Quantum Fourier Transformation and its inverse \\
$V$ &
Hilbert Space \\
$PV$ &
Present Value \\
$\oplus, \wedge$ &
\textsc{xor}, \textsc{and} \\
\end{tabular}
\newpage

\section{Quantum Computing Basics}
\label{sec:basics}
We will use the \textit{Quantum Circuit Model of Computation}. The model is described in a finite-dimensional complex \textit{Hilbert space} $V$. A \textit{quantum circuit} defines a sequence of operations which is applied to qubit registers that are initialized to a certain state. A \textit{qubit register} (also called \textit{qubit system}) is a list of qubits.\\

Before getting in the basics of quantum computing we introduce the useful \textit{Dirac Notation}. Vectors are written inside a "ket" $\ket{k}$, their dual is denoted with a "bra" $\bra{k}$. The scalar product of two vectors $\ket{k}$ and $\ket{l}$ is written as $\bra{l}\ket{k}$. This is why the notation is also called "bra-ket notation". Since  the Hilbert space is finite-dimensional, we can choose a fixed basis and enumerate the basis vectors instead of writing potentially large column vectors. The fixed basis is called \textit{computational basis} and can usually be associated with the canonical basis. Let for example $\CC^4$ be our vector space $V$. The computational basis may then be defined as
\[ \left\lbrace 
\begin{pmatrix}1\\0\\0\\0\end{pmatrix},
\begin{pmatrix}0\\1\\0\\0\end{pmatrix},
\begin{pmatrix}0\\0\\1\\0\end{pmatrix},
\begin{pmatrix}0\\0\\0\\1\end{pmatrix} 
\right\rbrace. \]
In Dirac notation we would enumerate the basis vectors, either in decimal or in binary notation, i.e.
\[ \{ \ket{0},\ket{1},\ket{2},\ket{3} \} \quad\textrm{or}\quad \{ \ket{00},\ket{01},\ket{10},\ket{11} \}, \]
which saves us a lot of writing in case of higher dimensions. The binary notation has the advantage, that the vectors can also be interpreted as tensor products, for example
\[ \ket{01} = \ket{0}\otimes\ket{1} = \begin{pmatrix}0\\1\\0\\0\end{pmatrix}. \]
We will often use a subscript to indicate the vector's size in terms of binary digits, i.e. the dimension of the vector space spanned by $\ket{k}_m, k=0,\dots,2^m-1,$ is $2^m$.\\

In the following sub-sections, we will briefly describe the four postulates of quantum mechanics: (\ref{sec:post1_state}) The state space postulate, (\ref{sec:post2_evo}) the evolution postulate, (\ref{sec:post3_measure}) the measurement postulate and (\ref{sec:post4_comp}) the composite systems postulate. We closely follow chapter 3 of \cite{kaye07}.

\subsection{A Quantum Bit}
\label{sec:post1_state}
\fbox{\begin{minipage}{\dimexpr\textwidth-2\fboxsep-4\fboxrule\relax}
\underline{State Space Postulate}\vspace{5pt}\\
The state of a qubit system is described by a unit vector in a Hilbert space $V$.
\end{minipage}}\\\\
As the smallest unit of computation, the \textit{Quantum Bit} or \textit{Qubit} is the quantum analogue of a bit on a classical computer. While the classical bit always is either in state $0$ or $1$, the qubit can be both at the same time. Mathematically, a qubit is a $2$-dimensional Hilbert space and all unit vectors in this space are potential states of the qubit. The state of a qubit $\ket{\psi}$ can hence be written as linear combination of the basis vectors, i.e. 
\[ \ket{\psi} = \alpha_0\ket{0} + \alpha_1\ket{1}, \]
where $\alpha_0,\alpha_1\in\CC$ with $|\alpha_0|^2+|\alpha_1|^2=1$ and $\{\ket{0},\ket{1}\}$ is an orthonormal basis. 
These linear combinations are called \textit{superpositions}. We can already see that the memory capacity of a qubit is much larger than of a classical bit: To represent the state of qubit we need to store two complex figures requiring each for example 1 byte = 8 bits, while a classical bit obviously needs 1 bit.\\

If we can write $e^{i\phi}\ket{\psi}$ for some $\phi\in\RR$, then $\phi$ is called \textit{global phase} of the qubit. Global phases are statistically irrelevant, so we can say that $e^{i\phi}\ket{\psi}$ and $\ket{\psi}$ represent the same state.\\

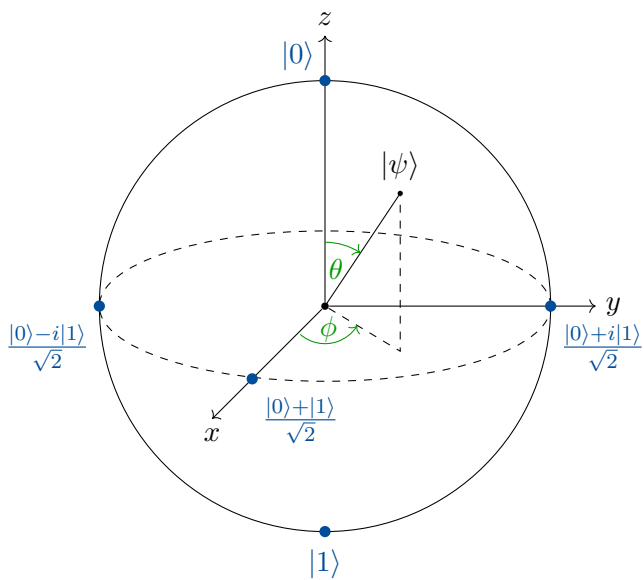
\begin{figure}[ht]
\begin{center}
\begin{tikzpicture}

    % Define radius
    \def\r{3}

    % Bloch vector
    \draw (0,0) node[circle,fill,inner sep=1] (orig) {} -- (\r/3,\r/2) node[circle,fill,inner sep=0.7,label=above:$\ket{\psi}$] (a) {};
    \draw[dashed] (orig) -- (\r/3,-\r/5) node (phi) {} -- (a);

    % Sphere
    \draw (orig) circle (\r);
    \draw[dashed] (orig) ellipse (\r{} and \r/3);

    % Axes
    \draw[->] (orig) -- ++(-\r/2,-\r/2) node[below] (x) {$x$};
    \draw[->] (orig) -- ++(\r*1.2,0) node[right] (y) {$y$};
    \draw[->] (orig) -- ++(0,\r*1.2) node[above] (z) {$z$};

    %Angles
    \pic [draw=dkgreen,text=dkgreen,->,"$\phi$"] {angle = x--orig--phi};
    \pic [draw=dkgreen,text=dkgreen,<-,"$\theta$",angle radius=.85cm] {angle = a--orig--z};
    
    %Poles
    \tkzDefPoint(0,\r){pz}
    \tkzDefPoint(\r,0){py}
    \tkzDefPoint(-\r/3.1,-\r/3.1){px}
    \tkzDefPoint(0,-\r){pmz}
    \tkzDefPoint(-\r,0){pmy}
        
    \tkzLabelPoint[above left,color=dkblue](pz){$\ket{0}$}
    \tkzLabelPoint[below right,yshift=-3,color=dkblue](py){$\frac{\ket{0}+i\ket{1}}{\sqrt{2}}$}
    \tkzLabelPoint[below right,yshift=-3,color=dkblue](px){$\frac{\ket{0}+\ket{1}}{\sqrt{2}}$}
    \tkzLabelPoint[below,yshift=-3,color=dkblue](pmz){$\ket{1}$}
    \tkzLabelPoint[below left,yshift=-3,color=dkblue](pmy){$\frac{\ket{0}-i\ket{1}}{\sqrt{2}}$}
    \foreach \n in {pz,py,px,pmz,pmy}
	  \node at (\n)[circle,fill,inner sep=1.5pt,color=dkblue]{};

\end{tikzpicture}
\end{center}
\caption{The Bloch sphere is a geometrical illustration of one qubit. All possible pure states of a qubit are located on its surface. The antipodal points correspond to the orthonormal basis and the equator represents the equally weighted superpositons between the basis vectors.}
\label{fig:bloch_sphere}
\end{figure}

A convenient and popular illustration of a qubit is the \textit{Bloch sphere} shown in figure \ref{fig:bloch_sphere}. All possible so called \textit{pure} states of a single qubit are located on its surface (see section \ref{sec:post4_comp} for the definition of pure vs. mixed states). The Bloch sphere representation of a qubit is based on polar coordinates:
\[ \ket{\psi} = \cos\left(\frac{\theta}{2}\right)\ket{0} + e^{i\phi}\sin\left(\frac{\theta}{2}\right)\ket{1}. \]
For $\theta=\pi$ we arrive at $e^{i\phi}\ket{1}\equiv\ket{1}$ and for a whole circle ($\theta=2\pi$) we get $-\ket{0}\equiv\ket{0}$. Hence dividing $\theta$ by $2$ is necessary to ensure that $2\pi$ corresponds to \textit{one} full circle. Finally note that the equally weighted superpositions of the basis states are located on the equator of the Bloch sphere.

\subsection{Evolution of a Quantum System}
\label{sec:post2_evo}
\fbox{\begin{minipage}{\dimexpr\textwidth-2\fboxsep-4\fboxrule\relax}
\underline{Evolution Postulate}\vspace{5pt}\\
Every evolution of a closed quantum system can be described by a \textit{unitary} transformation. That means that given an initial state $\ket{\psi_0}$ and a following state $\ket{\psi_1}$, there exists a \textit{linear operator} $\mathcal{U}:V\rightarrow V$ with $\mathcal{U}\mathcal{U}^\dagger=\mathcal{I}$ such that
\[ \ket{\psi_1} = \mathcal{U}\ket{\psi_0}. \]
\end{minipage}}\\\\
A linear operator is a linear transformation of a vector space to itself. The dagger ($\dagger$) denotes the Hermitian adjoint and $\mathcal{I}$ is the identity matrix.  Qubits in a closed system do not interact with qubits outside this system. In a circuit model, the operators are also called \textit{gates}. Note that the unitarity of quantum evolution implies that all quantum circuits are \textit{reversible}. Thus it is not possible to lose information during a quantum circuit since we can always "go back" to retrieve a previous state. Notable examples of quantum gates are:
\begin{itemize}
\item \textit{Pauli} gates are single qubit gates and perform a rotation by $\pi$ around the $x$- (Pauli-X), $y$- (Pauli-Y) or $z$-axis (Pauli-Z) of the Bloch sphere. For example Pauli-X is a logical \textsc{not}.
\item \textit{Square root} gates are "halved" Pauli gates, i.e. they do a quarter-turn in the Bloch sphere. Common names are S-gate for a rotation around the $z$-axis and SX-gate for the X-rotation, while the Y-rotation is rarely used. The square root notation is motivated from the fact that two successive applications deliver the repective Pauli gate, e.g. $\sqrt{Z}\sqrt{Z}\ket{\psi}=Z\ket{\psi}$.
\item The \textit{Hadamard} gate $X$ moves the basis vectors $\ket{0}$ and $\ket{1}$ to the equator of the Bloch sphere:
\begin{equation*}
\begin{aligned}
H\ket{0} &= \frac{1}{\sqrt{2}}(\ket{0}+\ket{1})\\
H\ket{1} &= \frac{1}{\sqrt{2}}(\ket{0}-\ket{1})
\end{aligned}
\end{equation*}
Hadamard gates play an important role in many quantum algorithms, for example in the amplitude estimation described below.
\item \textit{Controlled} gates are multi-qubit gates. A controlled $\mathcal{U}$ or \textsc{c}$\mathcal{U}$ applies the operator $\mathcal{U}$ to a \textit{target} register if the \textit{control} qubit is in state $\ket{1}$ and does nothing otherwise:
\begin{equation*}
\textsc{c}\mathcal{U}\ket{k}\ket{l}_m =
\begin{cases}
\ket{0}\ket{l}_m, & \ket{k}=\ket{0} \\
\ket{1}\mathcal{U}\ket{l}_m, & \ket{k}=\ket{1}.
\end{cases}
\end{equation*}
Controlled gates usually \textit{entangle} the control and the target qubits.
The \textit{controlled not} or \textsc{cnot} gate is probably the most important example, see section \ref{sec:post4_comp}.
\item \textit{Multi-controlled} gates have more than one control qubit. A multi-controlled operator is applied to a target register if all control qubits are in state $\ket{1}$ and does nothing otherwise:
\begin{equation*}
\textsc{c}^n\mathcal{U}\ket{k}_n\ket{l}_m =
\begin{cases}
\ket{k}_n\ket{l}_m, & \ket{k}_n\neq\ket{1\dots 1}_n \\
\ket{1\dots 1}_n\mathcal{U}\ket{l}_m, & \textrm{else}.
\end{cases}
\end{equation*}
An important example is the double-controlled \textsc{not} gate, also called $\textsc{ccnot}$ or \textit{Toffoli} gate.
\item The \textit{rotation} operators $R_x(\theta), R_y(\theta)$ and $R_z(\theta)$ rotate around the Bloch axes $x, y$ and $z$. They are generalizations of the Pauli gates.
\end{itemize}
Note that a linear operator is fully characterized by its action on a basis. Hence it suffices to define or analyze the operator on the basis states. The general case for an arbitrary superposition then follows from linearity. Furthermore the matrix representation of an operator is often not needed explicitly, since the case-by-case definition on the basis vectors is more convenient.

\subsection{Measurement}
\label{sec:post3_measure}
\fbox{\begin{minipage}{\dimexpr\textwidth-2\fboxsep-4\fboxrule\relax}
\underline{Measurement Postulate}\vspace{5pt}\\
For a given orthonormal basis $B=\{\ket{\varphi_k}\}$ and an arbitrary state 
\begin{equation}
\label{eq:measure_superpos}
\ket{\psi}=\sum_k\alpha_k\ket{\varphi_k},
\end{equation}
it is possible to perform a measurement with respect to $B$ which outputs label $k$ with probability $|\alpha_k|^2$ and leaves the system in state $\ket{\varphi_k}$.
\end{minipage}}\\\\
Thus, a superposition only exists until we perform a measurement. After a measurement, a qubit system is always in a basis state. Let for example $\ket{\psi}$ be in an equally weighted superposition of $\ket{0}$ and $\ket{1}$. We could say that that the qubit system represents a coin flip with unbiased coin. Before measurement, the coin is rotating and the state is $\ket{0}$ and $\ket{1}$. After measurement the state is $\ket{0}$ or $\ket{1}$, but we don't have much information about the probabilities. If we are interested in the probabilities, we could repeat the experiment several times and use an statistical estimator.\\

Furthermore note that we get $\alpha_k=\bra{\varphi_k}\ket{\psi}$ by applying the linear transformation $\bra{\varphi_k}$ to both sides of equation \eqref{eq:measure_superpos} and using the orthonormality. Hence the probability for obtaining result $k$ is the same given state $\ket{\psi}$ or $e^{i\phi}\ket{\psi}$, which shows that the global phase $e^{i\phi}$ is statistically irrelevant:
\begin{equation}
\label{eq:prob_alpha}
|\alpha_k|^2 = \bra{\psi}\ket{\varphi_k}\bra{\varphi_k}\ket{\psi}.
\end{equation}
This equation also shows a connection between projectors and measurement. An operator $\mathcal{P}$ is called \textit{projector} if $\mathcal{P}^2=\mathcal{P}$. For example $\mathcal{P}=\ket{\varphi_k}\bra{\varphi_k}$ projects a vector to the subspace spanned by $\ket{\varphi_k}$ and according to equation \eqref{eq:prob_alpha} it "extracts" the probability for output $k$ from $\ket{\psi}$. This statement can be generalized thanks to the \textit{Spectral Theorem}, which ensures the existence of the following decomposition for every \textit{normal} operator $\mathcal{U}$:
\[ \mathcal{U} = \sum_k \lambda_k\ket{\varphi_k}\bra{\varphi_k}, \]
where $\{\ket{\varphi_k}\}$ is an orthonormal basis consisting of eigenvectors and $\{\lambda_k\}$ are the corresponding eigenvalues. Normality ($\mathcal{P}^\dagger\mathcal{P}=\mathcal{P}\mathcal{P}^\dagger$) follows from unitarity. Hence the measurement postulate ensures that the following expression can be evaluated:
\begin{equation}
\label{eq:ev_observable}
\bra{\psi}\mathcal{U}\ket{\psi} = \sum_k \lambda_k\bra{\psi}\ket{\varphi_k}\bra{\varphi_k}\ket{\psi} = \sum_k \lambda_k |\alpha_k|^2.
\end{equation}
$\bra{\psi}\mathcal{U}\ket{\psi}$ is the expected value of $\mathcal{U}$'s eigenvalues given state $\ket{\psi}$. Finally, this leads us to the definition of observables: An operator $\mathcal{U}$ is called \textit{observable}, if it is Hermitian ($\mathcal{U}=\mathcal{U}^\dagger$). In this case all eigenvalues of $\mathcal{U}$ are real and hence the expected value \eqref{eq:ev_observable} is real as well. We could say, that given a qubit system $\ket{\psi}$ in a certain state, the observable determines which quantities are measured.

\subsection{Composite Systems}
\label{sec:post4_comp}
\fbox{\begin{minipage}{\dimexpr\textwidth-2\fboxsep-4\fboxrule\relax}
\underline{Composite System Postulate}\vspace{5pt}\\
If two qubit systems are combined, the state space of the composite system is the tensor product of the two corresponding Hilbert spaces. Given a state $\ket{\psi_0}$ from the first and a state $\ket{\psi_1}$ from the second system, the state in the composite system is
\[ \ket{\psi_0} \otimes \ket{\psi_1}. \]
\end{minipage}}\\\\
As seen during the introduction of Dirac notation, like for usual products, we often omit the $\otimes$ symbol and write $\ket{\psi_0}\ket{\psi_1}$ or $\ket{\psi_0\psi_1}$. When we assemble quantum circuits, we will often add qubit systems to an existing system. The composite system postulate tells us, that we can extend existing circuits and that the extended vector spaces are built by tensor products. Operators can be extended accordingly. If for example, we want to apply an operator $\mathcal{U}$ to the first system and do nothing on the second, we would write:
\[ (\mathcal{U} \otimes \mathcal{I})(\ket{\psi_0} \otimes \ket{\psi_1}) = \mathcal{U}\ket{\psi_0}\otimes\ket{\psi_1}. \]
Note that we won't omit the $\otimes$ between operators, because $\mathcal{U}\mathcal{I}(\ket{\psi_0} \otimes \ket{\psi_1})$ would suggest that $\mathcal{U}$ and $\mathcal{I}$ work each on the composite system.\\

Beside the ability of having two states at the same time, being \textit{entangled} with others is the second important feature of qubits. If two qubits are entangled, they interact with each other. Mathematically two qubits are entangled, if their state cannot be written as a product. Let for example $\ket{\psi}_2$ be in state:
\begin{equation}
\label{eq:epr_pair}
\ket{\psi}_2 = \textsc{cnot}(H\ket{0}\ket{0}) = \frac{1}{\sqrt{2}}\ket{0}\ket{0} + \frac{1}{\sqrt{2}}\ket{1}\ket{1},
\end{equation}
where $H$ is a Hadamard gate. Then the state of the first qubit is always equal to the second one. Hence it suffices to measure the state of one qubit to know the state of the entire system. If we actually measure the second qubit, the first qubit will be in state $\ket{0}$ or $\ket{1}$ with probability $\nicefrac{1}{2}$ each. Note that this is not a superposition. We have statistical information about the state and repeated measuring leads to the same result than measuring a qubit in the "corresponding" superposition. However, proceeding the circuit is different. Let for example $\ket{\psi}$ be in the entangled superposition \eqref{eq:epr_pair}. If we then measure the second qubit, we denote the state of the first qubit as:
\begin{equation}
\label{eq:mixed_state}
\ket{\psi} = \left\lbrace \left(\ket{0}, \frac{1}{2}\right), \left(\ket{1}, \frac{1}{2}\right) \right\rbrace .
\end{equation}   
Applying a Y-rotation $R_y$ by $\nicefrac{\pi}{2}$ the resulting state is
\begin{equation}
\ket{\psi} = \left\lbrace \left(H\ket{0}, \frac{1}{2}\right), \left(H\ket{1}, \frac{1}{2}\right) \right\rbrace ,
\end{equation}
i.e. the resulting state is $H\ket{0}$ or $H\ket{1}$ and measuring still outputs the same result (probability for $\ket{0}$ and $\ket{1}$ is $\nicefrac{1}{2}$ each). On the other hand, if we apply $R_y(\nicefrac{\pi}{2})$ to $H\ket{0}$ (which is statistically equal to \eqref{eq:mixed_state}), the result is $\ket{1}$. We call \eqref{eq:mixed_state} a \textit{mixed state} opposed to \textit{pure states}. Mixed states consist of one or more pure states. In the Bloch sphere visualization, mixed states are located within the sphere, while pure states lie on the surface. The mixed state \eqref{eq:mixed_state} represents the center of the Bloch sphere and hence rotations have no effect. We will use the technique of measuring parts of a system and proceeding with the rest in a mixed state during the amplitude estimation described in section \ref{sec:ae}.

\section{Building Blocks of Payoff Valuation on a Quantum Computer}
\label{sec:calc_e}
\begin{figure}[ht]
\centerline{
%\hspace{75pt}
\Qcircuit @C=1em @R=1em {
&&& \lstick{\ket{\rho}_m:} & \gate{\mathcal{H}} & \qw & \qw & \sgate{\mathcal{E}_2}{2} & \meter & \rstick{\hspace{100pt}}\\
&&& \lstick{\ket{\psi}_r\hspace{2pt}:} & \gate{\mathcal{A}} & \multigate{1}{\mathcal{B}} & \qw & \qw & \qw  \\   
&&& \lstick{\ket{\varphi}_s\hspace{2pt}:} & \qw & \ghost{\mathcal{B}} & \multigate{1}{\mathcal{E}_1} & \multigate{1}{\mathcal{E}_2} & \qw \\
&&& \lstick{\ket{\,\cdot\,}\hspace{7pt}:} & \qw & \qw & \ghost{\mathcal{E}_1} & \ghost{\mathcal{E}_2} & \qw & \rstick{\raisebox{2.2em}{\hspace{-12pt}$\ket{\chi}_{s+1}$ }} \\ 
&&&& 1 & 2 & 3\textrm{a} & 3\textrm{b} & 4 
\inputgroupv{3}{4}{1em}{1em}{\ket{\upsilon}_{s+1}\qquad}
\gategroup{3}{9}{4}{9}{1em}{\}}
}
}
\caption{General illustration of the building block assembling. The distribution loading (1) is followed by the payoff encoding (2). After that, the calculation of the expected value (3) is split into two sub-steps: $\mathcal{E}_1$ encodes the expected value to a single qubits amplitude (3a) and $\mathcal{E}_2$ followed by a measurement represents the amplitude estimation itself (3b+4). The intermediate state $\ket{\chi}_{s+1}$ plays an important role in the implementation of AE (see section \ref{sec:ae}). The gate $\mathcal{H}$ represents the initialization of the so called query register $\ket{\rho}_m$, which is needed for the amplitude estimation. Note that $\ket{\chi}_{s+1}$ denotes a certain state while $\ket{\upsilon}_{s+1}$ is the name of the qubit register. We will see that after step 3a, the register $\ket{\upsilon}_{s+1}$ is in state $\ket{\chi}_{s+1}$.}
\label{fig:general_bb}
\end{figure}
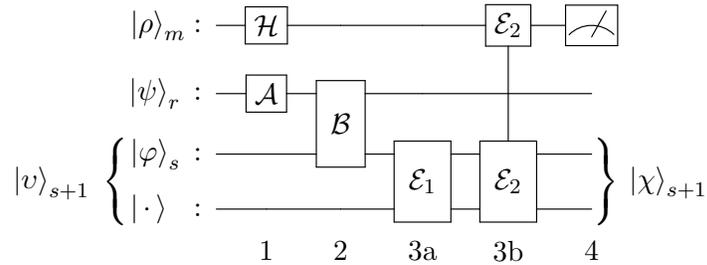

The valuation of a payoff essentially is the calculation of an expected value $\EE(Z)$ where $Z$ is the random variable representing the payoff behavior. The building blocks needed for modeling the valuation in a quantum circuit are (1) loading the market model, (2) implementing the payoff and (3) calculating the expected value. I.e. we start with loading the distribution of an underlying and proceed with encoding the payoff distribution based on this underlying. Given the payoff distribution, the expected value can finally be derived. Figure \ref{fig:general_bb} illustrates the general assembling. The crucial step for improving the Monte Carlo convergence is the calculation of the expected value. Doing this straightforward by measuring has no advantage over classical Monte Carlo, because the Monte Carlo error only decreases with $\sqrt{M}$ where $M$ is the number of \textit{shots} (executions of the circuit including measurement). The technique to achieve the quadratic speed-up is called \textit{Amplitude Estimation} (AE). To apply AE, we split the expected value calculation into two sub-steps: (3a) the encoding of the expected value to a single qubit's amplitude and (3b+4) the AE itself. As the main motivation for our work is the Monte Carlo speed-up via AE, we dedicate an own section to an "AE deep dive" after the building blocks (1)-(3) have been described.\\

For the general description of the building blocks, it suffices to consider $Z$ as univariate random variable, discretized with resolution $2^r$. The stochastic process interpretation will be introduced in connection with the insurance-related quantum circuits in section \ref{sec:quantum_circuits}.   

\subsection{Distribution Loading}
\label{sec:distr_loading}
Let $Z$ be a random variable and assume that we have a discretization of its range $[z_{min},z_{max}]$ to $2^r$ points $\{z_0,\dots,z_{2^r-1}\}$. The corresponding probabilities are $a_k:=\PP(Z=z_k), k=0,\dots,2^r-1$. We need a quantum algorithm $\mathcal{A}$ that prepares the state: 
\begin{equation}
\label{eq:distr_load}
\ket{\psi}_r = \mathcal{A}\ket{0}_r := \sum_{k=0}^{2^r-1}\sqrt{a_k}\ket{k}_r, 
\end{equation}
where the integer $k$ denotes the $k^{th}$ basis vector of the qubit system $\ket{\psi}_r$. Furthermore we define a mapping $k\rightarrow z_k$ as affine transformation: 
\begin{equation}
\label{eq:aff_trans_kx}
k \rightarrow z_{min} + \frac{z_{max}-z_{min}}{2^r-1}k.
\end{equation}
In other words the algorithm $\mathcal{A}$ maps an $r$-qubit register from state $\ket{0}_r$ to the entangled superposition $\ket{\psi}_r$ where each basis state $\ket{k}_r$ corresponds to a discretization point of $Z$. Hence we can use $\mathcal{A}$ as notation for the random variable $Z$ in quantum terms. See \cite{grover02} for a specific algorithm and \cite{iten16} for a generic approach. Zoufal et al. \cite{zoufal19} present a more efficient methodology of distribution loading using quantum Generative Adverserial Networks (qGANs).

\subsection{Payoff Implementation}
\label{sec:payoff_implementation}
In order to represent the payoff, the quantum circuit with the distribution register is extended by a second qubit register. Given the extended circuit in state $\ket{\psi}_{r}\otimes\ket{0}_{s}$, the next task is the construction of an algorithm $\mathcal{B}$, which encodes the payoff behavior into the second register:
\[ \ket{\psi}_{r}\otimes\ket{\varphi}_{s} := \mathcal{B}(\mathcal{A}\otimes\mathcal{I}_{s})\ket{0}_{r+s}. \]
After applying the payoff algorithm $\mathcal{B}$, the superposition $\ket{\varphi}_{s}$ represents the payoff's probability distribution. This is a key feature of quantum computing: Without explicit formulation of the distribution and without calculating each possible tuple of outcome $z$ and corresponding probability $\PP(Z=z)$ one by one, the quantum circuit provides the entire payoff behavior in a single step. The target register $\ket{\varphi}_{s}$ contains all information about the distribution at the same time. This is also remarkable in terms of memory: While a classical computer would need ca. $2^s$ bytes to represent a discrete probability distribution with $2^s$ possible values, a quantum computer only needs $s$ qubits. Of course the exact amount of classical memory required depends on the desired precision of the probabilities. We are assuming one byte (= 8 bits) per probability for this simple comparison but the exponential difference remains no matter which granularity we need. \\

The formal derivation of the circuits will often assume the involved qubit registers in basis states, e.g. $\ket{\psi}_r=\ket{k}_r$ in equation (\ref{eq:distr_load}). The general case with $\ket{\psi}_r$ in any superposition follows from linearity of all operators and the fact that a linear operator is completely defined by its action on a basis. Remember that all operators are linear according to the evolution postulate (\ref{sec:post2_evo}).

\subsection{Calculation of the Expected Value}
\label{sec:ev_calc}
Now that the payoff is represented by the state $\ket{\varphi}_{s}$, the last step is the calculation of the expected value. As described at the beginning of this section, this is where the main result of quantum computing in connection with Monte Carlo methods comes into play: The quadratic speed-up via amplitude estimation.\\ 

The first task represented by $\mathcal{E}_1$ in figure \ref{fig:general_bb} is to encode the expected value to a single qubit's amplitude, i.e. the initial state for the AE itself needs to be prepared. To avoid unnecessary complexity during the derivation of the AE algorithm, we assume that the payoff distribution is equal to the underlying distribution. I.e. step 2 in figure \ref{fig:general_bb} becomes needless and we can use $\ket{\psi}_r$ defined according to equation (\ref{eq:distr_load}) as starting point for step 3a. Once AE is implemented, it can be easily applied to arbitrary payoffs by plugging them in the circuit as algorithm $\mathcal{B}$.\\

Now, given step 1 completed, the register $\ket{\psi}_r$ contains full information about the discretized distribution and hence i.a. the expected value:
\begin{equation}
\label{eq:exp_value}
\EE[Z]:=\sum_{k=0}^{2^r-1}{a_k}z_k.
\end{equation}

The amplitude encoding is described in \cite{woerner19} with multi-controlled Y-rotations being the key components. We will describe some details of the implementation because it illustrates the current status of quantum algorithm development fairly well: Programming has to be done on low level (qubit level) and we deal a lot with mapping between binary representations, quantum amplitudes and real numbers.\\

\begin{figure}[ht]
\centerline{
\hspace{75pt}
\Qcircuit @C=1em @R=1em {
\lstick{\ket{k_0}} & \ctrl{5} & \qw & \qw & \qw & \qw & \qw & \qw \\ 
\lstick{\ket{k_1}} & \qw & \ctrl{4} & \qw & \qw & \qw & \qw & \qw\\
\lstick{\vdots} &&&& \ddots \\
&\\
\lstick{\ket{k_{r-1}}} & \qw & \qw & \qw & \qw & \qw & \ctrl{1} & \qw \\
\lstick{\ket{0}} & \gate{R_y(\vartheta_r)} & \gate{R_y(2\vartheta_r)} & \qw & \cdots && \gate{R_y(2^{r-1}\vartheta_r)} & \qw\\
}
}
\caption{Circuit to encode an integer on an amplitude. The total rotation performed on the last qubit depends on the binary representation of $k$. With $\vartheta_r=\nicefrac{2\pi}{2^r}$ being the rotation granularity, the total rotation can be uniquely mapped back to the integer value of $k$.}
\label{fig:circuit_laf}
\end{figure}
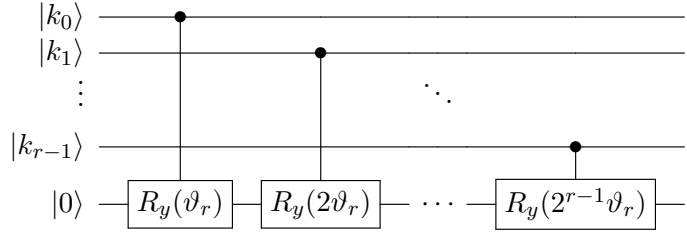

A Y-rotation is defined as
\[ R_y(\vartheta)\ket{0} := \cos(\frac{\vartheta}{2})\ket{0} + \sin(\frac{\vartheta}{2})\ket{1}. \]
The action on $\ket{1}$ follows from $R_y(\vartheta)\ket{1}=\textsc{not}(R_y(\vartheta)\ket{0})$. Now we assume the register $\ket{\psi}_r$ in a certain basis state $\ket{k}_r$ with binary representation $k=k_0\dots k_{r-1}$ and corresponding integer value $k\in\{0,\dots,2^r-1\}$. The full rotation ($\vartheta=2\pi$) is then split into $2^r$ parts and the integer $k$ determines how many of these parts are summed up to the realized rotation in random event $k$: 
\begin{equation}
\label{eq:theta_k}
\vartheta_k := \frac{2\pi k}{2^r} = \frac{\pi}{2^{r-1}} \sum_{i=0}^{r-1}2^i k_i \in \left[ 0,2\pi-\frac{2\pi}{2^r} \right]. 
\end{equation}
Note that $\{\vartheta_0,\dots,\vartheta_{2^r-1}\}$ is the discretization of the random variable $\theta$ which is an affine transformation of $Z$. Each term of the sum can be implemented as a controlled Y-rotation with the register $\ket{\psi}_r$ used as control qubits. The corresponding circuit is illustrated in figure \ref{fig:circuit_laf}.\\

After the rotation, the value of $k$ is encoded on the amplitude of the single ancilla qubit, i.e. we know how to implement the algorithm $\mathcal{E}_1$:
\[ \mathcal{E}_1\ket{k}_r\ket{0} = \ket{k}_r \left( \cos(\frac{\vartheta_k}{2})\ket{0} + \sin(\frac{\vartheta_k}{2})\ket{1} \right). \]
Please note that we only need single controlled rotations in this case. The mentioned multi-controlled rotations are needed if the approach is generalized to polynomials $f(k)=\sum_{i=0}^{n-1} c_ik^i$ instead of $f(k)=k$. \\

If we change $\ket{\psi}_r$ from a certain basis state to a general superposition, we get by linearity of $\mathcal{E}_1$:
\begin{equation}
\label{eq:ctrl_rot}
\begin{aligned}
\mathcal{E}_1\ket{\psi}_r\ket{0} & = \sum_{k=0}^{2^r-1}\sqrt{a_k} \mathcal{E}_1\ket{k}_r\ket{0} \\
& = \sum_{k=0}^{2^r-1}\sqrt{a_k} \ket{k}_r \left( \cos(\frac{\vartheta_k}{2})\ket{0} + \sin(\frac{\vartheta_k}{2})\ket{1} \right).
\end{aligned}
\end{equation}
Hence measuring the last qubit in state $\ket{1}$ gives us the probability:
\begin{equation}
\label{eq:def_p}
p := \sum_{k=0}^{2^r-1}a_k \sin^2\left(\frac{\vartheta_k}{2}\right)
\end{equation}
$p$ is formally obtained by a measurement of the observable $\mathcal{I}_{2^r}\otimes\ket{1}\bra{1}$ on the entangled state (\ref{eq:ctrl_rot}), where $\mathcal{I}_n$ is the $n$-dimensional identity matrix. More intuitively $p$ is the sum of the probabilities of the combined events "first qubit register in state $\ket{k}_r$ and second qubit in state $\ket{1}$". To get the expected value of $\vartheta$ (and hence $\EE[Z]$ by linearity of $\EE$, (\ref{eq:aff_trans_kx}) and (\ref{eq:theta_k})) the following linear approximation of $\sin^2$ is used:
\[ \sin^2\left(z+\frac{\pi}{4}\right) \approx z + \frac{1}{2} ~~\textrm{for}~~|z|<\frac{1}{2}.\]
Therefore we implement the affine transformation $\hat{\vartheta}_k/2:=(\vartheta_k/(2\pi)-1/2)c_{\textrm{approx}}+\pi/4$ with small $c_{\textrm{approx}}$ instead of $\vartheta_k$ in the rotation and get
\begin{equation*}
\hat{p} 
:= \sum_{k=0}^{2^r-1}a_k \sin^2\left(\frac{\hat{\vartheta}_k}{2}\right) 
\approx \sum_{k=0}^{2^r-1}a_k \left[ \left( \frac{\vartheta_k}{2\pi} - \frac{1}{2} \right) c_{\textrm{approx}}+\frac{1}{2} \right]
\end{equation*}
as measurement result from the circuit. That finally delivers an approximation for the transformed expected value:
\[ \EE\left(\frac{\vartheta}{2}\right) = \sum_{k=0}^{2^r-1}a_k \left(\frac{\vartheta_k}{2}\right) \approx \pi \left[ \left( \hat{p}-\frac{1}{2} \right)  \frac{1}{c_{\textrm{approx}}}+ \frac{1}{2} \right]. \] 
The transformations (\ref{eq:theta_k}) and (\ref{eq:aff_trans_kx}) finally lead us to the expected value $\EE(Z)$:
\[ \EE(Z) = z_{min} + \frac{z_{max}-z_{min}}{2^r-1} \EE\left(\frac{\vartheta}{2}\right) \frac{2^r}{\pi}.
\]
Implicitly assuming the existence of the approximation and the transformations, we will use the following loose notation for (\ref{eq:ctrl_rot}), which makes the encoding of the expected value to a single qubit's amplitude directly visible:
\begin{equation}
\label{eq:def_chi}
\ket{\chi}_{r+1} := \mathcal{E}_1\ket{\psi}_r\ket{0} = \sum_{k=0}^{2^r-1}\sqrt{a_k} \ket{z_k}_r \left( \sqrt{1-z_k}\ket{0} + \sqrt{z_k}\ket{1} \right).
\end{equation}
Now we know how to encode the expected value on the amplitude of a single qubit, i.e. how to prepare the state $\ket{\chi}_{r+1}$ from figure \ref{fig:general_bb}. Actually we did not encode the expected value itself, but a quantity $p$ from which we can derive the expected value by a linear transformation. Note that until now, there is no benefit compared to classical Monte Carlo. However, we prepared the ground for amplitude estimation, which will eventually deliver the expected value with the above-mentioned quadratic speed-up.\\

The amplitude estimation represented by $\mathcal{E}_2$ in figure \ref{fig:general_bb} followed by a measurement completes the abstract circuit. Therefore the circuit is extended by an additional qubit register $\ket{\rho}_m$, which is called \textit{query register}. We will give an in-depth description of AE in section \ref{sec:ae} so that we are able to assemble the concrete circuit afterwards (section \ref{sec:ae_summary}). 

\section{Amplitude Estimation}
\label{sec:ae}
The basic technique of amplitude estimation was developed by Brassard in 2002 \cite{brassard02} and is based on a generalization of Grover's search algorithm \cite{grover96}. Recent publications on this topic apply the idea to equity options \cite{rebent18}, show concrete implementations including real hardware results \cite{woerner19} and propose more efficient approaches which reduce the quantum computational effort \cite{suzuki20}. We will first describe the basic technique in section \ref{sec:ae_pe} and then give a short overview of recent developments.\\

Before we start describing the components of amplitude estimation, we clarify the vocabulary: 
\begin{itemize}
\item \textbf{Phase Estimation} (PE) helps finding a phase. If a qubit register is in state \eqref{eq:pep} then the task of PE is to find an estimation for the phase $x$. The \textit{Inverse Quantum Fourier Transformation} $QFT^{-1}$, introduced in section \ref{sec:pe} is an algorithm which solves the PE problem.
\item \textbf{Amplitude Amplification} (AA) algorithms amplify the amplitude of a sought basis state. This can for example be implemented by \textit{Phase Kick-Back} followed by PE (section \ref{sec:phase_kick_back}). A more basic example is the Grover search, where the sought answer is represented by a certain (but unknown) basis state and AA amplifies the probability for measuring this state.
\item \textbf{Amplitude Estimation} (AE) is a general term for circuits which deliver estimates of an amplitude, i.e. of a probability for measuring a certain state. The most basic example is repeated measurement. Section \ref{sec:ae_pe} presents a more sophisticated version where AA is used to amplify the probability of a state associated with the sought amplitude. I.e., in this case, the answer searched by AA is an amplitude itself.
\end{itemize} 

To connect the terms, we can say that the amplitude estimation algorithms described in this section are implemented via amplitude amplification. The latter can be performed with (section \ref{sec:ae_pe}) or without (section \ref{sec:ae_wope}) phase estimation.\\

In the following sections we will always treat cases, where the final outcome is the exact searched value. The reason why the algorithms are called phase and amplitude \textit{estimations} is, that we only get estimates in the general case. We will analyze the general case in section \ref{sec:effort_ae}.

\subsection{Amplitude Estimation based on Phase Estimation}
\label{sec:ae_pe}
The general idea of amplitude estimation based on phase estimation can be summarized as follows: Given a qubit register in state \eqref{eq:def_chi}, the probability for the event "last qubit = $\ket{1}$" is sought. Therefore a query register $\ket{\rho}_m$ is added to the circuit, whose basis states represent the possible results. Then the amplitude of the sought basis state (i.e. the one representing the expected value) is amplified. Thanks to the amplified amplitude, the final measurement returns the right answer with high probability so that ideally one execution of the circuit suffices. In other words we are exploiting the fact, that the quantum computer "knows" the answer without calculating all scenarios one-by-one. AE "extracts" the answer.\\

We will start with the formulation of the \textit{Phase Estimation} (PE) problem. The solution will naturally motivate the definition of the \textit{Quantum Fourier Transformation} (QFT), whose action on the enumerated basis states $\ket{l}\in\{\ket{0},\ket{1},\dots,\ket{2^{m-1}}\}$ is:
\begin{equation}
\label{eq:qft}
QFT_{2^m}\ket{l} := \frac{1}{\sqrt{2^m}}\sum_{k=0}^{2^m-1}e^{2\pi i\frac{l}{2^m} k}\ket{k}_m.
\end{equation}
Being able to implement PE, we will "kick" the expected value from the amplitude of $\ket{\chi}_{r+1}$ to the relative phase of a superposition $\ket{\rho}_m$ on which we can eventually apply PE.

\subsubsection{Quantum Phase Estimation}
\label{sec:pe}
The main result of this subsection will be the following: Let a qubit register $\ket{\rho}_m$ be in the superposition
\begin{equation}
\label{eq:pep}
\ket{\rho}_m = \frac{1}{\sqrt{2^m}}\sum_{k=0}^{2^m-1}e^{2\pi ixk}\ket{k}_m
\end{equation}
with $x\in[0,1]$. Then there is an \textit{efficient} quantum algorithm to obtain an estimate of $x$.\\

The task of estimating $x$ is called phase estimation problem. An algorithm is classified as efficient if the number of required resources (e.g. quantum gates) is bounded by an polynomial, i.e. it is in $\mathcal{O}(m^n)$ for a fixed $n\in\NN$.\\

In this section we will actually deal with the special case where $x=\frac{l}{2^m}$ for some $l\in\{0,\dots,2^m-1\}$, i.e. $\ket{\rho}_m=QFT_{2^m}\ket{l}$. The general case is addressed in section \ref{sec:effort_ae}. We denote the corresponding arc length by $\theta=2\pi x$. Furthermore $x_1,\dots,x_m\in\{0,1\}$ are the binary digits of $x=0.x_1\dots x_m$.\\ 

The solution of the PE problem has basically two ingredients: The phase rotation operator and the Hadamard gate. The former is defined on single qubits as follows for $j\in\NN$:
\begin{equation}
\begin{aligned}
R_j\ket{0} & :=  \ket{0} \\
R_j\ket{1} & :=  e^{\frac{2\pi i}{2^j}}\ket{1}. \\
\end{aligned}
\end{equation}
Hence the inverse phase rotation operator is:
\begin{equation}
R^{-1}_j\ket{1} :=  e^{-\frac{2\pi i}{2^j}}\ket{1}. \\
\end{equation}
Given this rotation we are able to "rotate off" digits from the \textit{relative phase} of the superposition $\frac{1}{\sqrt{2}}(\ket{0}+e^{2\pi ix}\ket{1})$. If the $j^{th}$ digit of $x$ is $1$ then $R^{-1}_j$ rotates it off.
\[
R^{-1}_j \left( \frac{\ket{0}+e^{2\pi i0.x_1\dots x_{j-1}1x_{j+1}\dots x_m}\ket{1}}{\sqrt{2}} \right) = \left( \frac{\ket{0}+e^{2\pi i0.x_1\dots x_{j-1}0x_{j+1}\dots x_m}\ket{1}}{\sqrt{2}} \right).
\]
If we focus on the last digit, the operator $R^{-1}_m$ rotates a point between $\frac{2\pi (k-1)}{2^{m-1}}$ and $\frac{2\pi k}{2^{m-1}}$ to the lower angle. Since the operator isn't useful if $x_j=0$, we will actually use controlled rotation operators when we assemble the circuit, i.e. the rotation is only applied if the corresponding digit is equal to $1$.\\

The Hadamard gate is self-inversive, i.e. $HH\ket{x_j}=\ket{x_j}$. Additionally, as $x_j\in\{0,1\}$, it can be written as: 
\[H\ket{x_j}=\frac{\ket{0}+(-1)^{x_j}\ket{1}}{\sqrt{2}}. \]

For $m=1$ and $x=0.x_1$, an application of Euler's formula shows that the state $\ket{\rho}$ defined in equation (\ref{eq:qft}) is equal to $H\ket{x_1}$:
\begin{equation}
\begin{aligned}
\ket{\rho} = \frac{1}{\sqrt{2}} \sum_{k=0}^{1}e^{2\pi i(0.x_1)k}\ket{k} & = \frac{\ket{0}+e^{\pi ix_1}\ket{1}}{\sqrt{2}} \\
& = \frac{\ket{0}+(\cos (\pi x_1) + i\sin (\pi x_1)\ket{1}}{\sqrt{2}} \\
& = \frac{\ket{0}+(-1)^{x_1}\ket{1}}{\sqrt{2}} \\
& = H\ket{x_1}.
\end{aligned}
\end{equation}

Due to the self-inversive property of $H$ we get:
\[ H\left( \frac{\ket{0}+e^{\pi ix_1}\ket{1}}{\sqrt{2}}  \right) = HH\ket{x_1} = x_1.\]

To see how the Hadamard gate eventually can help solving the phase estimation problem, we need the following identity:
\begin{equation}
\label{eq:qft_tensor}
\begin{aligned}
\frac{1}{\sqrt{2^m}} \sum_{k=0}^{2^m-1}e^{2\pi ix k}\ket{k}_m & = \bigotimes_{j=1}^m  \left( \frac{\ket{0}+e^{2\pi i (2^{m-j}x)}\ket{1}}{\sqrt{2}} \right) \\
& = \bigotimes_{j=1}^m \left( \frac{\ket{0}+e^{2\pi i (0.x_{m-j+1}x_{m-j+2}\dots)}\ket{1}}{\sqrt{2}} \right).
\end{aligned}
\end{equation}
The second equality follows from the fact that the multiplication $2^{m-j}x$ shifts the first $m-j$ digits of $x$ before the decimal separator and we can ignore full rotations because of $e^{2\pi ik}=1$ for $k\in\NN$.\\

Note that if $x=\frac{l}{2^m}$ for some integer $l$, the binary representation ends after $m$ digits. I.e. the relative phase of the first tensor factor (\ref{eq:qft_tensor}) has one digit ($0.x_m$), the second one has two digits ($0.x_{m-1}x_m$) and so on.\\

Now we are prepared to assemble the circuit for the inverse $QFT$. That proves the existence of an algorithm which maps the relative phase of a superposition to a basis state representing the phase: 
\begin{equation}
\textrm{QFT}_{2^m}^{-1}: \frac{1}{\sqrt{2^m}}\sum_{k=0}^{2^m-1}e^{2\pi i\frac{l}{2^m} k}\ket{k}_m \rightarrow \ket{l}_m.
\end{equation} 

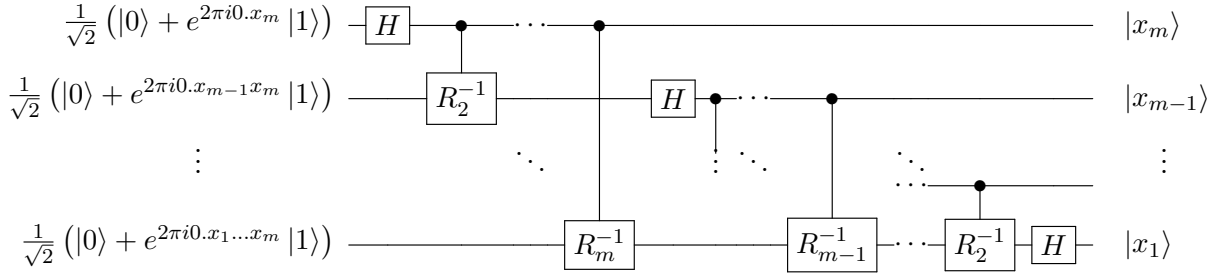
\begin{figure}[ht]
\centerline{
\Qcircuit @C=.59em @R=1em {
&&&&&&&&&&&& \lstick{ \frac{1}{\sqrt{2}}\left(\ket{0}+e^{2\pi i0.x_{m}}\ket{1} \right) } & \gate{H} & \ctrl{1} & \qw & \cdots & & \ctrl{4} & \qw & \qw & \qw & \qw & \qw & \qw & \qw & \qw & \qw & \qw & \qw & \qw & \rstick{ \ket{x_{m}} } \\
&&&&&&&&&&&& \lstick{ \frac{1}{\sqrt{2}}\left(\ket{0}+e^{2\pi i0.x_{m-1}x_{m}}\ket{1} \right) } & \qw & \gate{R^{-1}_2} & \qw & \qw & \qw & \qw & \gate{H} & \ctrl{1} & \qw & \cdots & & \ctrl{3} & \qw & \qw & \qw & \qw & \qw & \qw &\rstick{ \ket{x_{m-1}} } \\
&&&&&&&&&&&& \lstick{ \vdots\hspace{50pt} } & & & & \ddots & & & & \vdots & & \ddots & & & & \ddots & &  & & \rstick{ \hspace{20pt}\vdots } \\
&&&&&&&&&&&& & & & & & & & & & & & & & & \cdots & & \ctrl{1} & \qw & \qw & \\
&&&&&&&&&&&& \lstick{ \frac{1}{\sqrt{2}}\left(\ket{0}+e^{2\pi i0.x_{1}\dots x_{m}}\ket{1} \right) } & \qw & \qw & \qw & \qw & \qw & \gate{R^{-1}_m} & \qw & \qw & \qw & \qw & \qw & \gate{R^{-1}_{m-1}} & \qw & \cdots & & \gate{R^{-1}_{2}} & \gate{H} & \qw & \rstick{ \ket{x_1} } \\
}
}
\caption{Illustration of the Inverse Quantum Fourier Transformation. The last digit $x_m$ can be immediately determined by applying the Hadamard gate. For the other digits, in each case the higher digits have to be rotated off before application of the Hadamard gate. Hence in total $m$ Hadamard gates and $\frac{m(m-1)}{2}$ rotation operators are needed which, i.e. the QFT algorithm requires a total number of $\mathcal{O}(m^2)$ gates.}
\label{fig:qft}
\end{figure}

The inverse $QFT$ algorithm illustrated in figure \ref{fig:qft} can be described as follows:

\begin{enumerate}
\item Suppose a qubit register in state (\ref{eq:qft}) with $x=0.x_1x_2\dots x_m$ where $x_j\in\{0,1\}$ for $j=1,\dots,m$, written as tensor product according to (\ref{eq:qft_tensor})
\item For $j=1,\dots m$:
\begin{enumerate}
\item Apply Hadamard gate to the $j^{th}$ tensor factor: $H\left( \frac{\ket{0}+e^{2\pi i0.x_{m-j+1}}\ket{1}}{\sqrt{2}} \right) = \ket{x_{m-j+1}}$
\item For $i=1,\dots m-j$:
\begin{enumerate}
\item If $x_{m-j+1} = 1$: Apply inverse rotation operator $R^{-1}_{i+1}$ to $(i+1)^{th}$ tensor factor
\end{enumerate}
\end{enumerate}
\end{enumerate}

The crucial point is that if we are at step (2a) the relative phase of the current tensor factor has exactly one digit, because all following digits have been rotated off before. Please note that given a fault-tolerant quantum computer we don't run the circuit several times because the result register $\ket{x_1\dots x_m}_m$ is in a basis state and hence the measurement always returns the same result. The size $m$ of the qubit register determines the estimation accuracy as it corresponds to the number of binary digits which are calculated. The total number of gates required is $m$ Hadamard gates plus $\frac{m(m-1)}{2}$ rotation operators. Hence the resources needed for the $QFT$ algorithm is in $\mathcal{O}(m^2)$, i.e. the algorithm is efficient. Finally note that $QFT$ directly follows from the inverse $QFT$ by inverting all gates and running the circuit backwards. Moreover the order of the result register has to be reversed which can technically be achieved by the help of \textsc{swap} gates. \\

We have seen that for $x=\frac{l}{2^m}$ for some integer $l$, the algorithm delivers the exact phase value. In general, the result will be close to the exact value with high probability. We will discuss the estimation error and the computational effort in section \ref{sec:effort_ae}.

\subsubsection{Phase Kick-Back}
\label{sec:phase_kick_back}
We have introduced controlled operators in section \ref{sec:basics}. A controlled operator is applied to a tuple consisting of a control qubit and a target register. If the control qubit is in state $\ket{1}$, the operator is applied to the target, otherwise it has no effect. Hence it seems, that the control qubit remains unchanged after a controlled operation. However, the control qubit can also be affected due to \textit{phase kick-back}, where eigenstates and eigenvalues play the decisive role. Let $\mathcal{Q}$ be a $(1+2^m)$-dimensional operator with eigenstates $\ket{\psi_1}_m, \ket{\psi_2}_m$ and corresponding eigenvalues $\lambda_1, \lambda_2$. If we apply $\mathcal{Q}$ to an eigenstate we get $\mathcal{Q}\ket{\psi_i}_m=\lambda_i\ket{\psi_i}_m$. Hence, the controlled $\mathcal{Q}$ (we write \textsc{c}$\mathcal{Q}$) has the following effect if the target qubit is in an eigenstate:
\begin{equation}
\textsc{c}\mathcal{Q}\ket{k}\ket{\psi_i}_m =
\begin{cases}
\ket{0}\ket{\psi_i}_m, & \ket{k}=0 \\
\ket{1}\mathcal{Q}\ket{\psi_i}_m=\ket{1}\lambda_i\ket{\psi_i}_m=\lambda_i\ket{1}\ket{\psi_i}_m, & \ket{k}=1.
\end{cases}
\end{equation}
I.e. the target register in an eigenstate remains unchanged and the eigenvalue $\lambda_i$ can be associated with the control qubit: It is \textit{kicked-back} from the target to the control qubit. If the control qubit is in a superposition, we get by linearity:
\[ (\alpha_1\ket{0}+\alpha_2\ket{1})\mathcal{Q}\ket{\psi_i}_m=(\alpha_1\ket{0}+\alpha_2\lambda_i \ket{1})\ket{\psi_i}_m. \]
We see that the eigenvalue appears as relative phase of the control qubit. This effect will be used during amplitude estimation to "kick" the sought amplitude to the relative phases of control qubits where we eventually apply phase estimation.\\

Finally we have a look at the general case, where the target qubit is in a superposition of eigenstates. The controlled operator prepares an entangled state:
\begin{equation*}
\begin{aligned}
(\alpha_1\ket{0}+\alpha_2\ket{1})\mathcal{Q}(\beta_1\ket{\psi_1}_m+\beta_2\ket{\psi_2}_m) & = \beta_1(\alpha_1\ket{0}+\alpha_2\lambda_1 \ket{1})\ket{\psi_1}_m + \beta_2(\alpha_1\ket{0}+\alpha_2\lambda_2 \ket{1})\ket{\psi_2}_m) \\
& = \beta_1\ket{\varphi_1}\ket{\psi_1}_m + \beta_2\ket{\varphi_2}\ket{\psi_2}_m,
\end{aligned}
\end{equation*}
where $\ket{\varphi_i}=\alpha_1\ket{0}+\alpha_2\lambda_1 \ket{1}$. Due to the entanglement, we know that after measuring the second qubit the first register will be in state $\ket{\varphi_i}$ with probability $|\beta_i|^2$, i.e. in a mixed state (see section \ref{sec:post4_comp}). This result together with the following example will be important for the implementation of amplitude estimation.\\

\hrule\vspace{4pt}
\textbf{Example: Phase Kick-back + Phase Estimation = Amplitude Amplification} Let $\mathcal{Q}$ be a Y-rotation by $2\theta$ in the Bloch sphere. Then the matrix representation of $\mathcal{Q}$ is
\[ \mathcal{Q} = \begin{bmatrix}
\cos\theta & -\sin\theta \\
\sin\theta & \cos\theta
\end{bmatrix}. \]
The eigenvalues of $\mathcal{Q}$ are $e^{i\theta}$ and $e^{-i\theta}$ with the corresponding eigenstates denoted as $\ket{\psi_{\pm}}$. With the control being initialized by $H\ket{0}$ and the target qubit in an eigenstate $\ket{\psi_{\pm}}$ the controlled operator \textsc{c}$\mathcal{Q}$ prepares the following state:
\begin{equation}
\begin{aligned}
H\ket{0}\textsc{c}\mathcal{Q}\ket{\psi_{\pm}} & = \left(\frac{\ket{0}+\ket{1}}{\sqrt{2}} \right) e^{\mp i\theta} \ket{\psi_{\pm}} \\
& = \left(\frac{\ket{0}+e^{\mp i\theta}\ket{1}}{\sqrt{2}} \right) \ket{\psi_{\pm}}.
\end{aligned}
\end{equation} 
Note that for $\theta=2\pi 0.x$ with $x\in\{0,1\}$, we have shown in section \ref{sec:pe} that $x$ can be determined by applying an Hadamard gate (i.e. performing phase estimation for $m=1$) to the first qubit:
\[
H \left(\frac{\ket{0}+e^{\mp 2\pi i0.x}\ket{1}}{\sqrt{2}} \right) = \ket{x}.
\]
\hrule\vspace{12pt}

This is a basic example of \textit{Amplitude Amplification}. The initial state of the control qubit is an equally weighted superposition of the basis states $\ket{0}$ \textit{and} $\ket{1}$. After phase kick-back and phase estimation, the state is $\ket{x}$, i.e. either $\ket{0}$ \textit{or} $\ket{1}$ depending on the rotation angle. That means that the amplitude of $\ket{x}$ has been amplified from $\frac{1}{\sqrt{2}}$ to $1$.\\ 

\begin{figure}[ht]
\centerline{
\Qcircuit @C=.65em @R=1em {
\lstick{ \ket{0} } & \gate{H} & \ctrl{5} & \qw & \qw & \qw & \qw & \qw & \qw & \multigate{4}{QFT^{-1}} & \qw & \rstick { \ket{x_m} } \\
\lstick{ \ket{0} } & \gate{H} & \qw & \ctrl{4} & \qw & \qw & \qw & \qw & \qw & \ghost{QFT^{-1}} & \qw & \rstick { \ket{x_{m-1}} } \\
\lstick{ \vdots } & & & & & \ddots \\
& \\
\lstick{ \ket{0} } & \gate{H} & \qw & \qw & \qw & \qw & \qw & \ctrl{1} & \qw & \ghost{QFT^{-1}} & \qw & \rstick { \ket{x_1} }\\
\lstick{ \ket{\psi_{\pm} } } & \qw & \gate{\mathcal{Q}^{2^{m-1}}} & \gate{\mathcal{Q}^{2^{m-2}}} & \qw & \cdots && \gate{\mathcal{Q}} & \qw & \qw & \qw & \rstick{\ket{\psi_{\pm} }}\\
}
}
\caption{Illustration of an amplitude amplification circuit. The control register is initialized with equal weights and the target qubit is initialized with an eigenstate $\ket{\psi_{\pm} }$ of the rotation. The latter remains unchanged. After kicking back the eigenvalues, the phase estimation algorithm $QFT^{-1}$ is applied. Eventually the amplitude of the basis state $\ket{2^mx}$ with $x=\nicefrac{\theta}{2\pi}$ is equal to $1$ while all other amplitudes are $0$.}
\label{fig:aa_pe}
\end{figure}
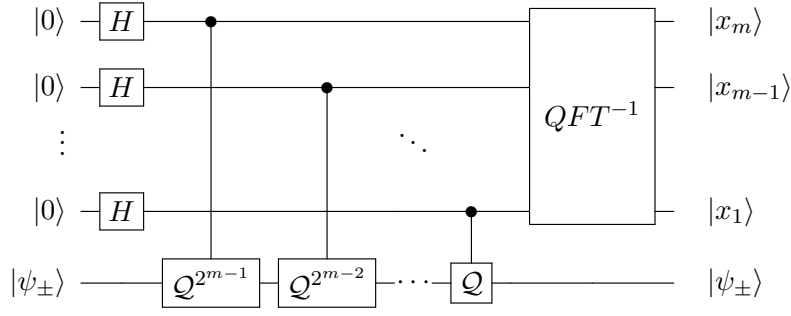

For the general case with $\theta=2\pi 0.x_1\dots x_m$ for $x_j\in\{0,1\}$ we are using the fact that $e^{j\pm i\theta}$ are eigenvalues of repeated rotations $\mathcal{Q}^j$. Hence we can prepare the state $\frac{1}{\sqrt{2}}\left( \ket{0}+e^{\mp 2\pi i0.x_{j+1}\dots x_m}\ket{1} \right)$ by applying $\mathcal{Q}^{2^j}$ to the target qubit. That means that given a Y-rotation we can encode the rotation angle to a qubit register in a way that this qubit register is in the initial state for the phase estimation. We eventually have built an algorithm for amplifying the amplitude of the basis state which represents the rotation angle. The corresponding circuit is illustrated in figure \ref{fig:aa_pe}. \\
\newpage

Note that the initialization of the query register is achieved by an $m$-qubit Hadamard transformation $H^{\otimes m}$, which results in an equally weighted superposition. As $H^{\otimes m}=QFT\ket{0}_m$ we can directly see that the circuit works in the most simple case: $\theta=0\implies \mathcal{Q}=\mathcal{I}\implies \ket{x}_m=QFT^{-1}QFT\ket{0}_m=\ket{0}_m$.\\

For the sake of comprehensiveness, we amend some details on the eigenstates: The eigenstates of the Y-rotation are $\ket{\psi_{\pm}}=\frac{1}{\sqrt{2}}\left(\ket{\chi}\pm i\ket{\chi^{\bot}}\right)$, where $\ket{\chi}$ and $\ket{\chi^{\bot}}$ are orthonormal vectors which span the $xz$-subspace in the Bloch sphere. In other words $\ket{\chi}$ and $\ket{\chi^{\bot}}$ span the subspace which is mapped to itself by the Y-rotation. The eigenstate property follows straightforward from applying $\mathcal{Q}$, e.g. for $\ket{\psi_+}$ we get:
\begin{equation*}
\begin{aligned}
\mathcal{Q}\ket{\psi_+} & = \frac{1}{\sqrt{2}}\left( \mathcal{Q}\ket{\chi} + i \mathcal{Q}\ket{\chi^{\bot}} \right)\\
& = \frac{1}{\sqrt{2}} \left[\left(\cos\theta\ket{\chi}+\sin\theta\ket{\chi^{\bot}}\right)+i\left(-\sin\theta\ket{\chi}+\cos\theta\ket{\chi^{\bot}}\right)\right] \\
& = \frac{1}{\sqrt{2}} \left[\left(\cos\theta-i\sin\theta\right)\ket{\chi}+\left(\sin\theta+i\cos\theta\right)\ket{\chi^{\bot}}\right] \\
& = e^{-i\theta}\frac{1}{\sqrt{2}}\left[\ket{\chi}+i\ket{\chi^{\bot}}\right] \\
& = e^{-i\theta}\ket{\psi_+}.
\end{aligned}
\end{equation*}

\subsubsection{Application to Expected Value Calculation}
\label{sec:app_ev}
The preceding sections have shown how we can implement amplitude estimation given a rotation operator. The remaining task is to find the "suitable" rotation in the subspace spanned by $\ket{\chi}_{r+1}$ and $\ket{\chi^{\bot}}_{r+1}$ where $\ket{\chi}_{r+1}$ is defined by equation (\ref{eq:def_chi}). We will denote this subspace by $H(\chi,\phi)$. Suitable means that the rotation angle must be related to the expected value. Once we found this rotation we can WLOG assume that $\ket{\chi}_{r+1}$ is a linear combination of the rotation's eigenstates, apply amplitude estimation and eventually derive the expected value from the estimated phase. In other words, the task is to define a rotation without knowing the angle. We will closely follow \cite{rebent18} in the following derivation. To ease notation, we will omit the subscript of $\ket{\chi}_{r+1}$ \\

First we are looking for a function $f$ which defines a relationship between the expected value encoded in the amplitude of $\ket{\chi}$ and an arbitrary rotation angle in the hyperplane $H(\chi,\phi)$. Let $\mathcal{V}$ be a linear operator defined as follows:
\[ \mathcal{V}:=\mathcal{I}_{2^{r+1}}-2\mathcal{I}_{2^r}\otimes\ket{1}\bra{1}. \]
Note that $\mathcal{V}$ is a slight modification from the definition in section \ref{sec:calc_e} to make it a unitary operator, i.e. $\mathcal{V}^{-1}=\mathcal{V^\dagger}$. Thanks to the measurement postulate we know that we can measure the unitary $\mathcal{V}$ on $\ket{\chi}$, which delivers a linear transformation of the expected value:   
\begin{equation}
\label{eq:def_mu}
\bra{\chi}\mathcal{V}\ket{\chi} = 1-2\mu,
\end{equation}
where $\mu$ is defined according to equation \eqref{eq:def_p}, i.e.
\begin{equation}
\label{eq:def_mu2}
\mu := \sum_{k=0}^{2^r-1}a_k \sin^2\left(\frac{\vartheta_k}{2}\right).
\end{equation}
Since we know how the expected value can be derived from $\mu$ (see section \ref{sec:calc_e}) we can use $\mu$ as synonym for the expected value.\\

As $\mathcal{V}$ is a unitary operator, $\mathcal{V}\ket{\chi}$ can be written as a Bloch vector using the orthonormal basis $\{\ket{\chi},\ket{\chi^{\bot}}\}$:
\[ \mathcal{V}\ket{\chi} = \cos\left(\frac{\theta}{2}\right)\ket{\chi} + e^{i\phi}\sin\left(\frac{\theta}{2}\right)\ket{\chi^{\bot}}. \]
This representation leads us to $\bra{\chi}\mathcal{V}\ket{\chi}=\cos\left(\nicefrac{\theta}{2}\right)$ and hence it specifies a relationship between the expected value and an angle $\theta$ in the hyperplane $H(\chi,\phi)$. The searched function $f$ can be defined based on the identity $1-2\mu=\cos\nicefrac{\theta}{2}$:
\[ f(\theta):=\frac{1}{2}\left(1-\cos\left(\frac{\theta}{2}\right)\right)=\mu. \]

The remaining task of implementing a rotation in $H(\chi,\phi)$ with an unknown angle is solved by a sequence of reflections. Thereby the unknown angle must be linked to $\theta$ defined by $\mathcal{V}$. The result will actually be a rotation by $2\theta$. Given an arbitrary orthonormal basis $\{\ket{\varphi},\ket{\varphi^{\bot}}\}$, a unitary operator $\mathcal{U}$ is called \textit{reflection across} $\ket{\varphi^{\bot}}$ if:
\begin{equation}
\begin{aligned}
\mathcal{U}\ket{\varphi} &= -\ket{\varphi} \\
\mathcal{U}\ket{\varphi^{\bot}} &= \ket{\varphi^{\bot}}. \\
\end{aligned}
\end{equation}
A reflection can be implemented by $\mathcal{U}=\mathcal{I}-2\ket{\varphi}\bra{\varphi}$. Note that $-\mathcal{U}$ reflects across $\ket{\varphi}$.\\

The concrete reflections needed for our task are:
\begin{equation*}
\begin{aligned}
\mathcal{U} & = \mathcal{I}_{2^{r+1}}-2\ket{\chi}\bra{\chi} \\
\mathcal{S} & = \mathcal{I}_{2^{r+1}}-2\mathcal{V}\ket{\chi}\bra{\chi}\mathcal{V}. \\
\end{aligned}
\end{equation*} 
Note that both $\mathcal{U}$ and $\mathcal{S}$ are defined in the hyperplane $H(\chi,\phi)$.\\

\begin{figure}[ht]
\begin{center}
\begin{tikzpicture}

    % Define radius
    \def\r{3}

    % Bloch vectors
    \draw[line width=1.25pt,dkred] (0,0) node[circle,fill,inner sep=1] (orig) {} -- (0.5*\r,0.87*\r) node[circle,fill,inner sep=1,label=above:$\quad\ket{\upsilon}$] (a) {};
    \draw[line width=1.25pt,dkgray] (0,0) node[circle,fill,inner sep=1] (orig) {} -- (0.75*\r,0.66*\r) node[circle,fill,inner sep=1,label=right:$\mathcal{V}\ket{\upsilon}$] (b) {};
    \draw[line width=1.25pt,purple] (0,0) node[circle,fill,inner sep=1] (orig) {} -- (0.92*\r,0.39*\r) node[circle,fill,inner sep=1,label=right:$-\mathcal{S}\ket{\upsilon}$] (c) {};
    \draw[line width=1.25pt,dkgreen] (0,0) node[circle,fill,inner sep=1] (orig) {} -- (-0.92*\r,0.39*\r) node[circle,fill,inner sep=1,label=left:$\mathcal{Q}\ket{\upsilon}$] (d) {};    

    % Sphere
    \draw (orig) circle (\r);
    \draw[dashed] (orig) ellipse (\r{} and \r/3);

    % Axes
    \draw[->] (orig) -- ++(\r*1.2,0) node[right] (y) {};
    \draw[->] (orig) -- ++(0,\r*1.2) node[above] (z) {$z$};

    %Angles
    \pic [draw=dkred,text=dkred,<-,"$\gamma$",angle radius=.95cm,angle eccentricity=1.2,pic text options={scale=.8}] {angle = a--orig--z};
    \pic [draw=purple,text=purple,<-,"$2\theta-\gamma$",angle radius=1.95cm,pic text options={shift={(41pt,7pt)},scale=.8}] {angle = c--orig--z};
    \pic [draw=dkgreen,text=dkgreen,->,"-$2\theta$",angle radius=2.45cm,pic text options={shift={(-10pt,20pt)},scale=.8}] {angle = a--orig--d}; 
    \pic [draw=dkgray,text=dkgray,<-,"$\theta$",angle radius=1.45cm,pic text options={shift={(2pt,22pt)},scale=.8}] {angle = b--orig--z};
    %Poles
    \tkzDefPoint(0,\r){pz}
    \tkzDefPoint(\r,0){py}
    \tkzDefPoint(0,-\r){pmz}

    \tkzLabelPoint[above left,color=dkblue](pz){$\ket{\chi}$}
    \tkzLabelPoint[below right,yshift=-3,color=dkblue](py){$\frac{\ket{\chi}+e^{i\phi}\ket{\chi}^\bot}{\sqrt{2}}$}
    \tkzLabelPoint[below,yshift=-3,color=dkblue](pmz){$\ket{\chi}^\bot$}
  
    \foreach \n in {pz,py,pmz}
	  \node at (\n)[circle,fill,inner sep=1.5pt,color=dkblue]{};

\end{tikzpicture}
\end{center}
\caption{Illustration constructing the rotation operator $\mathcal{Q}$ in the Bloch sphere. The orthonormal basis vectors are $\ket{\chi}$ and $\ket{\chi}^\bot$ and the Z-rotation is fixed to an arbitrary angle $\phi$. Starting with the initial state $\ket{\upsilon}$ (red, angle=$\gamma$), a reflection across $\mathcal{V}\ket{\upsilon}$ leads to $-\mathcal{S}\ket{\upsilon}$ (purple, angle=$\gamma+2(\theta-\gamma)=2\theta-\gamma$. The following projection across $\ket{\chi}$ eventually leads to $\mathcal{Q}\ket{\upsilon}$ (green, angle=$-(2\theta-\gamma)$ and hence the angle between the initial and the final state is $-(2\theta-\gamma)+(-\gamma)=-2\theta$.}
\label{fig:rotations}
\end{figure}
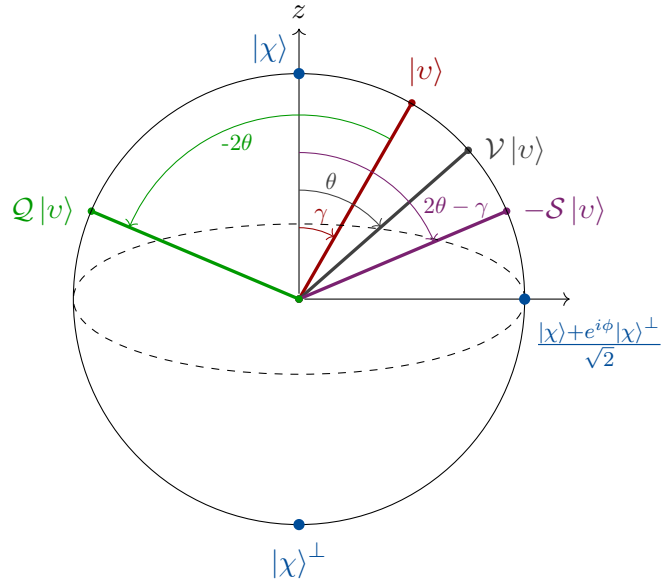

Now suppose that $\ket{\upsilon}$ is an arbitrary state in the hyperplane $H(\chi,\phi)$:
\[ \ket{\upsilon} = \cos\left(\frac{\gamma}{2}\right)\ket{\chi} + e^{i\phi}\sin\left(\frac{\gamma}{2}\right)\ket{\chi^{\bot}}, \]
where $\gamma$ is an arbitrary angle. Then, reflecting $\ket{\upsilon}$ across $\mathcal{V}\ket{\chi}$ leads to the intermediate state $-\mathcal{S}\ket{\upsilon}$:
\begin{equation*}
\begin{aligned}
-\mathcal{S}\ket{\upsilon} &= -\ket{\upsilon} + 2\mathcal{V}\ket{\chi}\left[\bra{\chi}\mathcal{V}\cos\left(\frac{\gamma}{2}\right)\ket{\chi} + \bra{\chi}\mathcal{V}e^{i\phi}\sin\left(\frac{\gamma}{2}\right)\ket{\chi^\bot} \right] \\
& = -\ket{\upsilon} + 2\mathcal{V}\ket{\chi}\left[\bra{\chi}\mathcal{V}\cos\left(\frac{\gamma}{2}\right)\ket{\chi} + e^{-i\phi}\sin\left(\frac{\theta}{2}\right)\bra{\chi^\bot}e^{i\phi}\sin\left(\frac{\gamma}{2}\right)\ket{\chi^\bot} \right] \\
& = -\ket{\upsilon} + 2\mathcal{V}\ket{\chi}\left[\cos\left(\frac{\gamma}{2}\right)\cos\left(\frac{\theta}{2}\right) + \sin\left(\frac{\theta}{2}\right)\sin\left(\frac{\gamma}{2}\right) \right] \\
& = \cdots \\
& = \cos\left(\frac{2\theta-\gamma}{2}\right)\ket{\chi} + e^{i\phi}\sin\left(\frac{2\theta-\gamma}{2}\right)\ket{\chi^{\bot}},
\end{aligned}
\end{equation*}
i.e. the angle of the intermediate state $-\mathcal{S}\ket{\upsilon}$ in the hyperplane $H(\chi,\phi)$ is $2\theta-\gamma$. This can be seen more easily in figure \ref{fig:rotations}. The formal derivation uses $\bra{\chi}\mathcal{V}=(\mathcal{V}\ket{\chi})^\dagger$ in step 2 and several trigonometric calculation rules in step 4.\\

\newpage
The next operation is a reflection of $-\mathcal{S}\ket{\upsilon}$ across $\ket{\chi}$, which effectively changes the sign of the angle. The rotation is implemented by an application of $-\mathcal{U}$. With $\mathcal{Q}:=-\mathcal{U}(-\mathcal{S})=\mathcal{U}\mathcal{S}$:
\begin{equation*}
\begin{aligned}
\mathcal{Q}\ket{\upsilon} &= \mathcal{S}\ket{\upsilon} + 2\ket{\chi}\left[ \cos\left(\frac{2\theta-\gamma}{2}\right)\ket{\chi} + e^{i\phi}\sin\left(\frac{2\theta-\gamma}{2}\right)\ket{\chi^{\bot}} \right] \\
&= \mathcal{S}\ket{\upsilon} + 2\cos\left(\frac{2\theta-\gamma}{2}\right)\ket{\chi} \\
&= \cos\left(-\frac{2\theta-\gamma}{2}\right)\ket{\chi} + e^{i\phi}\sin\left(-\frac{2\theta-\gamma}{2}\right)\ket{\chi^{\bot}}.
\end{aligned}
\end{equation*}

If we now look at the angle between the initial state $\ket{\upsilon}$ and the final state $\mathcal{Q}\ket{\upsilon}$ we see that its $-(2\theta-\gamma)+(-\gamma)=-2\theta$. As the two states are located in the same hyperplane, we found that the composed reflections $\mathcal{U}\mathcal{S}$ build a rotation about $-2\theta$. For the actual implementation, the reflections have to be broken down so that we can work in the computational basis. With $\mathcal{F}:=\mathcal{E}_1(\mathcal{A}\otimes\mathcal{I}_2)$ we already have implemented an algorithm to prepare $\ket{\chi}$:
\[ \ket{\chi} = \mathcal{F}\ket{0}_{r+1}. \]
Let $\mathcal{Z}$ be the reflection in the computational basis:
\[ \mathcal{Z} = \mathcal{I}_{2^{r+1}}-2\ket{0}_{r+1}\bra{0}_{r+1}. \]
As $\mathcal{F}$ is a unitary operator it holds $\mathcal{F}^{-1}=\mathcal{F}^\dagger$. Given that we can successively compose $\mathcal{U}$, $\mathcal{S}$ and finally $\mathcal{Q}$ by $\mathcal{F}, \mathcal{Z}$ and $\mathcal{V}$:
\begin{equation*}
\begin{aligned}
\mathcal{U} & = \mathcal{F}\mathcal{Z}\mathcal{F^\dagger} \\
\mathcal{S} & = \mathcal{V}\mathcal{U}\mathcal{V} = \mathcal{V}\mathcal{F}\mathcal{Z}\mathcal{F^\dagger}\mathcal{V}\\
\mathcal{Q} & = \mathcal{U}\mathcal{S} = \mathcal{F}\mathcal{Z}\mathcal{F^\dagger}\mathcal{V}\mathcal{F}\mathcal{Z}\mathcal{F^\dagger}\mathcal{V}. \\
\end{aligned}
\end{equation*}
The amplitude amplification algorithm $\mathcal{E}_2$ from figure \ref{fig:general_bb} can eventually be described as follows:
\[ \mathcal{E}_2: H^{\otimes m}\ket{0}_m\ket{\chi}_{r+1} \rightarrow \left(QFT^{-1}\otimes\mathcal{I}_{r+1}\right)\left(\bigotimes_{j=1}^{m-1}\textsc{c}\mathcal{Q}_j^{2^{m-j}}\right) (H^{\otimes m}\ket{0}_m\ket{\chi}_{r+1}     ), \]
where the controlled rotation operator $\mathcal{Q}_j^i$ uses the $j^{th}$ qubit of the first register as control and is applied $i$ times on the target register:
\begin{equation*}
\textsc{c}\mathcal{Q}_j^i: \ket{k}_m\ket{\chi}_{r+1} \rightarrow 
\begin{cases}
\ket{k}_m\mathcal{Q}^i\ket{\chi}_{r+1} , & \ket{k_j}=\ket{1} \\
\ket{k}_m\ket{\chi}_{r+1} , & \textrm{else}. \\
\end{cases}
\end{equation*} 

Now all components for implementing the quantum circuit which calculates the expected value of a random variable are prepared and we will assemble them in the following section.

\subsubsection{Assembling of the Quantum Circuit}
\label{sec:ae_summary}
\begin{figure}[ht]
\centerline{
\Qcircuit @C=1.em @R=1em {
&&&&&& \mbox{Amplitude Amplification } \mathcal{E}_2  \\
&&\lstick{ \ket{0} } & \multigate{4}{\mathcal{H}} & \qw & \ctrl{5} & \qw & \qw & \qw & \qw & \qw & \qw & \multigate{4}{QFT^{-1}} & \qw & \meter & \rstick { \ket{x_m} } \\
&&\lstick{ \ket{0} } & \ghost{\mathcal{H}} & \qw & \qw & \ctrl{4} & \qw & \qw & \qw & \qw & \qw & \ghost{QFT^{-1}} & \qw & \meter & \rstick { \ket{x_{m-1}} } \\
&&\lstick{ \vdots } & & & & & & \ddots \\
& \\
&&\lstick{ \ket{0} } & \ghost{\mathcal{H}} & \qw & \qw & \qw & \qw & \qw & \qw & \ctrl{1} & \qw & \ghost{QFT^{-1}} & \qw & \meter & \rstick { \ket{x_1} }\\
&&\lstick{ \ket{0} } & \multigate{3}{\mathcal{A}} & \multigate{4}{\mathcal{E}_1} & \multigate{4}{\mathcal{Q}^{2^{m-1}}} & \multigate{4}{\mathcal{Q}^{2^{m-2}}} & \qw & \cdots && \multigate{4}{\mathcal{Q}} & \qw & \qw & \qw & \qw \\
&&\lstick{ \vdots } & & & & & & \cdots \\
&&& & & & & & & & & & & & & \rstick{ \ket{\chi}_{r+1}} \\
&&\lstick{ \ket{0} } & \ghost{\mathcal{A}} & \ghost{\mathcal{E}_1} & \ghost{\mathcal{Q}^{2^{m-1}}} & \ghost{\mathcal{Q}^{2^{m-2}}} & \qw & \cdots && \ghost{\mathcal{Q}} & \qw & \qw & \qw & \qw \\
&&\lstick{ \ket{0} } & \qw & \ghost{\mathcal{E}_1} & \ghost{\mathcal{Q}^{2^{m-1}}} & \ghost{\mathcal{Q}^{2^{m-2}}} & \qw & \cdots && \ghost{\mathcal{Q}} & \qw & \qw & \qw & \qw 
\gategroup{2}{6}{11}{13}{.8em}{--}
\gategroup{7}{15}{11}{15}{1.2em}{\}}
\inputgroupv{2}{6}{1em}{3.5em}{\vspace{50pt}\ket{\rho}_m\quad}
\inputgroupv{7}{10}{1em}{2em}{\ket{\psi}_r\quad}\\
}
}
\caption{Illustration of the quantum circuit which calculated the expected value of a random variable by using amplitude amplification. The query register $\ket{\rho}_m$ is initialized by an $m$-qubit Hadamard gate and the state $\ket{\chi}_{r+1}$ is prepared by encoding the expected value to the amplitude of the last qubit. Afterwards $\mathcal{E}_2$ amplifies the amplitude of the basis state representing the expected value in the query register. The final measurement returns an estimation value $2^mx\in\NN$ which can be mapped to the expected value.}
\label{fig:ae}
\end{figure}
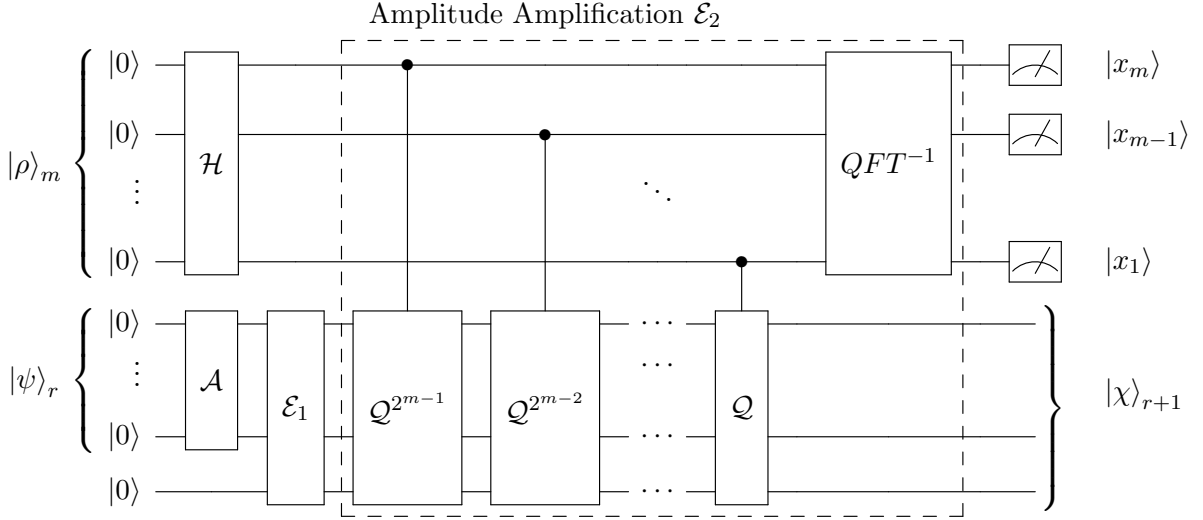

The abstract quantum circuit for calculating the expected value of a random variable using amplitude estimation is illustrated in figure \ref{fig:general_bb}. In the preceding section we have shown how the isolated building blocks look like and now we can finally put everything together. The resulting amplitude estimation circuit is illustrated in figure \ref{fig:ae}.

\begin{enumerate}
\item \textit{Distribution Loading} (section \ref{sec:distr_loading}) encodes a probability distribution to a qubit register:
\[ \mathcal{A}: \ket{0}_r \rightarrow \sum_{k=0}^{2^r-1}\sqrt{a_k}\ket{k}_r. \]
\item \textit{Payoff Implementation} (section \ref{sec:payoff_implementation}) transforms the probability distribution of an underlying to the probability distribution of a payoff:
\[ \mathcal{B}: \mathcal{A}\ket{0}_r\otimes\ket{0}_s \rightarrow \sum_{k=0}^{2^{(r+s)-1}}\sqrt{a_k}\ket{k}_r. \]
While the development of specific payoffs is our main contribution (see following sections), we omit this part of the circuit in this section in order to concentrate on the general mechanism of amplitude estimation. In other words, we assume that the payoff is equal to the underlying.
\item \textit{Calculation of the Expected Value} 
\begin{enumerate}
\item \textit{Encoding the Expected Value to an Amplitude} (section \ref{sec:ev_calc}) is achieved by controlled rotations.
\[ \mathcal{E}_1: \mathcal{A}\ket{0}_r\otimes\ket{0} \rightarrow \sum_{k=0}^{2^r-1}\sqrt{a_k} \ket{z_k}_r \left( \sqrt{1-z_k}\ket{0} + \sqrt{z_k}\ket{1} \right) := \ket{\chi}_{r+1}. \]
\item \textit{Amplitude Amplification} (section \ref{sec:ae}) is composed of the phase estimation algorithm (section \ref{sec:pe}) and the controlled rotation operator $\mathcal{Q}$ (section \ref{sec:app_ev}). 
\begin{equation*}
\begin{aligned}
\mathcal{E}_2: \ket{\chi}_{r+1}\otimes H^{\otimes m}\ket{0}_m & \rightarrow \left(QFT^{-1}\otimes\mathcal{I}_{r+1}\right)\left(\bigotimes_{j=1}^{m-1}\textsc{c}\mathcal{Q}_j^{2^{m-j}}\right) H^{\otimes m}\ket{0}_m\ket{\chi}_{r+1}\\
& = \ket{\chi}_{r+1}\otimes\ket{2^mx}_m,
\end{aligned}
\end{equation*}
where $x$ can be mapped to the expected value. The initialization of the query register is achieved by an $m$-qubit Hadamard transformation:
\[ \mathcal{H}: \ket{0}_m \rightarrow H^{\otimes m}\ket{0}_m = \frac{1}{\sqrt{2^m}}\sum_{k=0}^{2^m-1}\ket{k}. \]
\end{enumerate}
\item \textit{Measurement} completes the amplitude estimation. In an ideal setup with a fault-tolerant quantum computer and $x=\nicefrac{l}{2^m}$ for some integer $l$ the circuit has to be executed only once. In general, the number of shot depends on the required accuracy of the estimation.
\end{enumerate}

\subsubsection{Computational Effort compared to Classical Monte Carlo}
\label{sec:effort_ae}
In order to compare Monte Carlo estimators, we need to define the comparison criteria. Suppose that 
\[ \hat{\mu} = \frac{1}{n}\sum_{i=1}^M Z_i \]
with $Z_i$ iid, $\EE[Z_i]=\mu$ and $\VV [Z_i]=\sigma^2<\infty$, is a classical Monte Carlo estimator for the expected value. Then the central limit theorem leads to a relationship between the estimation error and the number of samples:
\[ \mu-\hat{\mu} \approx N\left(0,\frac{\sigma^2}{M}\right). \] 
This means that the estimation error is approximately normally distributed with variance $\nicefrac{\sigma^2}{M}$. Hence the following equation gives us the confidence interval for a designated confidence level $\alpha$:
\[ \PP\left[|\mu-\hat{\mu}|\leq q_{\alpha/2}\frac{\sigma}{\sqrt{M}}\right] \approx 1-\alpha, \]
where $q_{\alpha/2}=\Phi^{-1}(1-\nicefrac{\alpha}{2})$ and $\Phi$ is the standard normal distribution function. Hence the size of the error is in $\mathcal{O}(1/\sqrt{M})$. In other words, if we want to increase the accuracy by a factor $a$, the number of samples has to be increased by $a^2$ \cite{glass03}.\\

As quantum computers work fundamentally different, the comparison is not straightforward. For example in the amplitude estimation algorithm there is nothing directly comparable to the classical samples because the quantum computer simulates the entire distribution at the same time. Therefore we are introducing the term \textit{Quantum Samples} which denotes the number of possible basis states, which is $M=2^m$ for $m$ qubits. Quantum samples can be seen as analogue to classical Monte Carlo scenarios. Then we will assess the effort for running the circuit once, depending on the number of quantum samples and finally analyze the relationship between this effort and the estimation error.\\   

Assessing the effort for running the circuit actually means counting operators. The complexity of the amplitude estimation algorithm is dominated by the $2^m-1$ applications of $\mathcal{Q}$ (remember that QFT requires $\mathcal{O}(m^2)$ gates). Hence there is a linear relationship between the computational effort and the number of quantum samples $M$.\\     

Concerning the estimation error, we have already seen that it is zero if $x=\frac{l}{2^m}$ for some integer $l$. In this case we can say that $x$ fits the granularity $\frac{1}{2^m}$ of the query register $\ket{\rho}_m$. If $x$ is located between two discretization points, i.e. $\frac{k}{2^m}\leq x\leq\frac{k+1}{2^m}$, we can show that the phase estimation algorithm returns one of the two closest points $\frac{k}{\sqrt{2^m}}$ and $\frac{k+1}{\sqrt{2^m}}$ with probability at least $\frac{8}{\pi^2}$ (see below for the proof). Hence we found a relationship between the number of qubits of the query register $m$ and the estimation error:
\begin{equation}
\label{eq:est_error}
\PP\left[|x-\hat{x}|\leq \frac{1}{2^m}\right] \geq \frac{8}{\pi^2},
\end{equation}
where $\hat{x}\in[0,1]$ is the outcome of the AE circuit. This is the main result of the error estimation. We still need to translate the inaccuracy of $\hat{x}$ to the inaccuracy of the expected value. Actually it can be shown that if $|x-\hat{x}|\leq \frac{1}{2^m}$ then $|\mu-\hat{\mu}|\leq \mathcal{O}(\frac{1}{2^m})$ and hence with $M=2^m$:
\[ \PP\left[|\mu-\hat{\mu}|\leq \mathcal{O}\left(\frac{1}{M}\right)\right] \geq \frac{8}{\pi^2}, \]
where the definition of $\hat{\mu}$ is based on \eqref{eq:def_mu}:
\[ \hat{\mu}=\frac{1}{2}\left(1-\cos\frac{\hat{\theta}}{2}\right) \]
and $\hat{\theta}=2\pi\hat{x}$ is the transformed output from AE. In other words, the convergence rate improves from $\frac{1}{\sqrt{M}}$ for classical Monte Carlo to $\frac{1}{M}$ for amplitude estimation, which is a quadratic speed-up. We refer to \cite{brassard02} for the details.\\

We conclude this paragraph with the proof of equation \eqref{eq:est_error}. Let $x\in[0,1]$ be some fixed number. Then in the AE circuit, after application of the controlled rotations $\mathcal{Q}^j$, we have the following state in the query register:
\[ \ket{\rho}_m = \frac{1}{\sqrt{2^m}}\sum_{k=0}^{2^m-1}e^{2\pi ixk}\ket{k}_m. \]
Since $|x-\hat{x}|\leq\frac{1}{2^m}$ we can decompose $x$ in binary notation ($x_i\in\{0,1\}$):
\begin{equation}
\label{eq:decomp_x}
x = \hat{x} + \varepsilon_x = 0.x_1\dots x_m + 0.0\dots 0x_{m+1}\dots
\end{equation}
with $\varepsilon_x<\frac{1}{2^m}$. The following inverse QFT leads us to:
\begin{equation*}
\begin{aligned}
QFT_{2^m}^{-1}\ket{\rho}_m &= \frac{1}{\sqrt{2^m}}\sum_{k=0}^{2^m-1}e^{2\pi ixk}\frac{1}{\sqrt{2^m}}\sum_{l=0}^{2^m-1}e^{-2\pi i\frac{k}{2^m}l}\ket{l}_m \\
&= \frac{1}{2^m}\sum_{k=0}^{2^m-1}\sum_{l=0}^{2^m-1}e^{2\pi ixk}e^{-2\pi i\frac{k}{2^m}l}\ket{l}_m\\
&= \frac{1}{2^m}\sum_{l=0}^{2^m-1}\sum_{k=0}^{2^m-1}e^{2\pi i(\hat{x}+\varepsilon_x)k}e^{-2\pi i\frac{k}{2^m}l}\ket{l}_m\\
&= \frac{1}{2^m}\sum_{l=0}^{2^m-1}\sum_{k=0}^{2^m-1}e^{2\pi ik\left(\frac{2^m\hat{x}-l}{2^m}+\varepsilon_x\right)} \ket{l}_m\\
&=:  \frac{1}{2^m}\sum_{l=0}^{2^m-1}\alpha_{x,m}(l)\ket{l}_m
\end{aligned}
\end{equation*}
Hence computing the geometric sum delivers the amplitude of the state $\ket{l}_m=\ket{2^m\hat{x}}_m$:
\[ \alpha_{x,m}(2^m\hat{x}) = \frac{1}{2^m}\sum_{k=0}^{2^m-1}e^{2\pi ik\varepsilon_x} = \frac{1}{2^m}\left(\frac{1-e^{\pi i\varepsilon_x2^m}}{1-e^{\pi i\varepsilon_x}} \right), \] 
which leads us to the corresponding probability:
\[ \PP(\hat{x}) = |\alpha_x(2^m\hat{x})|^2 = \frac{1}{2^{2m}}\left|\frac{1-e^{\pi i\varepsilon_x2^m}}{1-e^{\pi i\varepsilon_x}} \right|^2 = \frac{1}{M^2}\left|\frac{\sin^2(M\pi\varepsilon_x)}{\sin^2(\pi\varepsilon_x)} \right|^2, \]
where the last step follows from $|1-e^{i2\theta}|=|e^{-i\theta}-e^{i\theta}|=2|\sin\theta|$. For $|\pi\varepsilon_x|\leq\nicefrac{\pi}{2}$ (which is ensured by \eqref{eq:decomp_x}) and for $M>1$, one can show that $\PP(\hat{x})\geq\nicefrac{4}{\pi^2}$ and since this is true for both neighbors of $x$, the probability for $|x-\hat{x}|\leq\nicefrac{1}{2^m}$ is at least $2\PP(\hat{x})=\nicefrac{8}{\pi^2}$.

\subsection{Amplitude Estimation without Phase Estimation}
\label{sec:ae_wope}
The conventional approach for amplitude estimation described in the previous section is difficult to implement on near-term quantum computers. A measure for the near-term implementability is the number of controlled operators, or on lowest level the number of \textsc{cnot} gates. Suzuki et al. proposed the \textit{Maximum Likelihood Amplitude Estimation} \cite{suzuki20}, Grinko et al. introduced \textit{Iterative Amplitude Estimation} \cite{grinko19}. Both variants do not rely on phase estimation and are hence less expensive in terms of \textsc{cnot}s. We won't analyze these approaches further in this work, but would like to emphasize that there are ongoing developments which bring us closer to the practical application of quantum computers.

\subsection{Excursus: Grover's Quantum Search Algorithm}
\label{sec:grover_search}
At the beginning of section \ref{sec:ae} we stated that the amplitude estimation is based on a generalization of Grover's search algorithm. To close the loop we will give a brief description of the connection between these algorithms. The search problem solved by Grover's algorithm is defined as follows: Given a black box $U_f$ for computing an unknown function $f:\{0,1\}^m\rightarrow \{0,1\}$, find an input $\{0,1\}^m$ such that $f(x)=1$. Thus it is easy to check if a solution is correct (one evaluation of $U_f$), but it is hard to find it ($2^m$ possible solutions). Grover eventually finds an operator, which amplifies the amplitude of the solution in an query register. Similar to the amplitude estimation described above, the query register starts in an equally weighted superposition and then successively moves towards the amplified state. In particular, after $k$ applications of the operator $G$ called \textit{Grover iterate}, the query register is in state:
\[ G^k H^{\otimes m}\ket{0}_m = \cos \left(\frac{(2k+1)\theta}{2}\right)\ket{\varphi}_m + \sin \left( \frac{(2k+1)\theta}{2}\right)\ket{\varphi^\bot}_m, \]
where $\ket{\varphi^\bot}_m$ is the searched solution and $\theta$ is defined by $\sin\nicefrac{\theta}{2}=\nicefrac{1}{\sqrt{2^m}}$. Hence if $(2k+1)\theta$ approaches $\pi$ the amplitude of the searched solution approaches $1$. In other words, the Grover iterate needs to be applied approximately $\nicefrac{\pi}{2\theta}-\nicefrac{1}{2}$ times to get $\ket{\varphi^\bot}_m$ with maximal probability.\\

Within the amplitude estimation algorithm the operator $\mathcal{Q}$ takes the role of the Grover iterate. In fact equation \eqref{eq:ctrl_rot} can be written as the following superposition:
\begin{equation*}
\begin{aligned}
\mathcal{E}_1\ket{\psi}_r\ket{0} &= \sum_{k=0}^{2^r-1}\sqrt{a_k} \ket{k}_r \left( \cos(\frac{\theta_k}{2})\ket{0} + \sin(\frac{\theta_k}{2})\ket{1} \right) \\
&= \cos\left(\frac{\theta}{2}\right)\ket{\psi_0}_r\ket{0} + \sin\left(\frac{\theta}{2}\right)\ket{\psi_1}_r\ket{1}
\\
&= \cos\left(\frac{\theta}{2}\right)\ket{\varphi}_{r+1} + \sin\left(\frac{\theta}{2}\right)\ket{\varphi^\bot}_{r+1}
\end{aligned}
\end{equation*}
with $\ket{\varphi}_{r+1}$ and $\ket{\varphi^\bot}_{r+1}$ located in the hyperplane $H(\chi,\phi)$. Since $\mathcal{Q}$ is a rotation about $2\theta$ we get:
\begin{equation}
\label{eq:mlh}
\mathcal{Q}^k \mathcal{E}_1\ket{\psi}_r\ket{0} =   \cos \left(\frac{(2k+1)\theta}{2}\right)\ket{\varphi}_{r+1} + \sin \left( \frac{(2k+1)\theta}{2}\right)\ket{\varphi^\bot}_{r+1}.
\end{equation}
This shows the connection between Grover's search and the amplitude estimation algorithm. Furthermore equation \eqref{eq:mlh} is used in the maximum likelihood amplitude estimation \cite{suzuki20} by measuring $\mathcal{Q}^k \mathcal{E}_1$ for $k=2^0,\dots,2^{m-1}$ to get an estimate for $\theta$.\\  

A detailed analysis of the Grover search is given in \cite{kaye07}, section 8.1.

\section{Insurance-related Payoffs}
\label{sec:payoff}
We will strongly abstract some general features of insurance contracts in this section. Our target is to motivate the further research on modeling such features to accelerate the growth of quantum circuit libraries. The ultimate goal is to enable actuaries to assemble their simulations from a toolkit of encapsulated quantum gates. This is obviously a long-term target. In the context of the IBM road map \cite{gambetta22}, actuaries should be put in the position to work as model developers and we want to contribute to the necessary basis of algorithms. Several contributions have already been made in the recent years, i.e. we don't start from zero. For example Stamatopoulos  et al. \cite{stama20} present a \textit{weighted adder}, which we utilize and illustrate in section \ref{sec:circ_whole_life}. They also propose a circuit which encodes the payoff of a barrier option. Therefore  \textit{comparator circuits} are introduced to check for barrier crossing in every time step.

\subsection{General Payoff}
\label{sec:general_payoff}
Let $Z=(Z_{t_i},i=1,\dots,n)$ be a discrete-time stochastic process and $\tau$ a stopping time (w.r.t. the filtration generated by $Z$). We will investigate different specifications of the random variable
\begin{equation} \label{eq:1}
Z_{\tau} := \sum_{i=1}^{n} \indic_{\{\tau=t_i\}}Z_{t_i}
\end{equation}
and calculate the corresponding expected values $PV:=\EE[Z_{\tau}]$. \\ 

In this work we will always interpret $Z_{t_i}$ as zero coupon bond prices, i.e. as discount factors. Hence $Z$ represents the stochastic process of interest rates. For the sake of simplicity we additionally assume all potential payments to be equal to $1$ so that these amounts do not explicitly appear in the formulas. For example the expected value of the present value of a certain payment (of $1$ unit) after three time steps would be $\EE[Z_{t_3}]$. The random variable $Z_{\tau}$ should be interpreted as the present value of a payment of $1$ unit at a uncertain time given stochastic discount factors. The uncertainty of the payment times is modeled via the stopping time $\tau$.\\ 

To model the quantum algorithms we furthermore assume discretized random variables $Z_{t_i}$:
\[Z_{t_i}:\Omega\rightarrow R_{t_i}, R_{t_i}:=\{z_{t_i,0},\dots,z_{t_i,2^{r_{t_i}}-1}\}\subseteq\RR, r_{t_i}\in\NN, \]
where we call $r_{t_i}$ the resolutions of the discretizations. In order to ease notation equal resolutions are assumed for all $Z_{t_i}$ and we set $r_{t_i}\equiv\hat{r}$. With $R^n:=R_{t_1}\times\dots\times R_{t_n}$ and $\textbf{z}=(z_{t_1},\dots,z_{t_n})\in R^n$, the expected value can eventually be written as
\begin{equation} \label{eq:2}
\begin{split}
PV & = \EE\left[\sum_{i=1}^{n} \indic_{\{\tau=t_i\}}Z_{t_i}\right] \\
& = \sum_{i=1}^{n} \sum_{\textbf{z}\in R^n} \PP  \left( \{\tau=t_i\} \cap \{Z=\textbf{z}\} \right) \cdot z_{t_i} \\
& = \sum_{i=1}^{n} \sum_{\textbf{z}\in R^n} \PP  \left( \{\tau=t_i\} | \{Z=\textbf{z}\} \right) \cdot \PP (Z=\textbf{z}) \cdot z_{t_i} .
\end{split}
\end{equation}
In the following section we will specify different characteristics of the stopping time $\tau$.

\subsection{Specifications}
\label{sec:specifications}
\subsubsection{Whole life insurance}
\label{sec:whole_life_intr}
The whole life insurance (or death-benefit insurance) pays a certain amount ($1$ unit in our case) in the event of death. With $Z_{t_i}$ interpreted as discount factors with duration $t_i$ and $\tau$ as residual lifetime, equation \eqref{eq:2} denotes the expected cash outflow of the insurance company. As the residual life time does not depend on interest rates, the stopping probabilities simplify to
\[ \PP  \left( \{\tau=t_i\} | \{Z=\textbf{z}\} \right) = \PP \left( \{T_x=t_i\} \right) = \prescript{}{t_i-1}{p_x} \cdot \prescript{}{t_i}{q_x} =: w^x_{t_i} \]
where we use the usual actuarial notation:
\begin{equation} \label{eq:3}
\begin{aligned}
\tau & := T_x  &  \textrm{residual lifetime of an x-year-old} \\
q_x & := \PP(T_x=1)  & \textrm{1-year mortality rate of an x-year-old} \\
p_x & := 1-q_x  & \textrm{1-year survival rate of an x-year-old} \\
\prescript{}{t_i}{q_x} & := \PP(T_x\leq t_i) & t_i\textrm{-year mortality rate of an x-year-old}\\
\prescript{}{t_i}{p_x} & := 1-\prescript{}{t_i}{q_x} & t_i\textrm{-year survival rate of an x-year-old}.
\end{aligned}
\end{equation}
The expected value $PV$ simplifies to the expected value of a weighted sum of random variables, where the weights are the probabilities of a residual life time equal to $t_i$ and the corresponding random variables are discount factors with duration $t_i$:
\begin{equation} \label{eq:4}
\begin{aligned}
PV & = \sum_{i=1}^{n} w^x_{t_i} \sum_{\textbf{z}\in R^n} \PP(\{Z=\textbf{z}\}) \cdot z_{t_i} \\
& =  \sum_{i=1}^{n} w^x_{t_i} \EE \left[ Z_{t_i} \right] \\
& =  \EE \left[ \sum_{i=1}^{n} w^x_{t_i} Z_{t_i} \right].
\end{aligned}
\end{equation}
Instead of calculating $n$ expected values of $Z_{t_i}$ one by one, the weighted sum can be implemented directly in a quantum circuit (see \ref{sec:circ_whole_life}).

\subsubsection{Dynamic Lapse}
\label{sec:dyn_lapse}
In usual life insurance contracts the policy holder has the right to cancel his contract and get a predefined payment. If the lapse rate depends on an underlying (e.g. on interest rates) this is called \textit{dynamic lapse}. With the predefined payment set to $1$ and $Z$ again interpreted as discount factors, the random variable \eqref{eq:1} describes the corresponding model. Therefore the stopping time $\tau$ is composed as follows:
\begin{equation} \label{eq:5}
\begin{aligned}
p_{t_i}: R_{t_i}\rightarrow\left[0,1\right], i=1,\dots,n & & \textrm{lapse rates at time }t_i \\
p(Z) = (p_{t_i}(Z_{t_i}), i=1,\dots,n) & &  \textrm{stochastic process of lapse rates} \\
L = (L_{t_i}, i=1,\dots,n), L_{t_i}\sim\mathrm{Bernoulli}(p_{t_i}(Z_{t_i}))  & &  \textrm{marginal lapse events} \\
\tau := \min_{i=1,\dots,n}\{L_{t_i}=1\} & & \textrm{stopping time}.
\end{aligned}
\end{equation}

To ensure a payment in case of no lapse event before $t_n$ we set $p_{t_n} = 1$. Given that, the lapse probabilities modeled by the described stopping time are:
\[ \PP  \left( \{\tau={t_i}\} | \{Z=\textbf{z}\} \right) = \prod_{j=1}^{i-1} \left( 1-p_{t_j}(z_{t_j}) \right) \cdot p_{t_i}(z_{t_i}). \]
I.e. the probability for contract cancellation at $t_i$ corresponds to the probability of surviving time steps $t_1$ until $t_{i-1}$ times the marginal lapse probability of time step $t_i$. This implies path dependency and ensures that a contract can only be canceled once.

\section{Insurance-related Quantum Circuits}
\label{sec:quantum_circuits}
As insurance contracts are often modeled over more than one period, we start this section with the introduction of stochastic processes. Section \ref{sec:distr_loading} describes the distribution loading for a general random variable $Z$. This random variable can also be interpreted as stochastic process $(Z_{t_i}, i=1,\dots,n)$. In this case, every basis state of the register represents a trajectory of the process. We define a partition $r_{t_1},\dots,r_{t_n}$ with $\sum_{i=1}^n r_{t_i}=r$ of the qubit register and interpret each part $i$ as marginal distribution of $Z_{t_i}$. To ease up notation we assume identical resolutions $\hat{r}$ in every time step and omit the sub-index $i$. Then we can write
\[ \ket{\psi}_r := \sum_{k_1,\dots,k_n} \sqrt{a_{k_1,\dots,k_n}}\ket{k_1}_{\hat{r}}\otimes\dots\otimes\ket{k_n}_{\hat{r}} \] 
with $a_{k_1,\dots,k_n}$ being the probability of the corresponding trajectory.  

\subsection{Whole life insurance}
\label{sec:circ_whole_life}
\begin{figure}[ht]
\centerline{
\Qcircuit @C=1em @R=1em {
\lstick{\psi_{t_1} \; \ket{0}_{\hat{r}}} & \multigate{4}{\mathcal{A}} & \ctrl{1} & \qw \\
\lstick{\psi_{t_2} \; \ket{0}_{\hat{r}}} & \ghost{\mathcal{A}} & \ctrl{1} & \qw \\
\lstick{\psi_{t_3} \; \ket{0}_{\hat{r}}} & \ghost{\mathcal{A}} & \ctrl{2} & \qw \\
\vdots & \nghost{\mathcal{A}} & & \\
\lstick{\psi_{t_n} \; \ket{0}_{\hat{r}}} & \ghost{\mathcal{A}} & \ctrl{1} & \qw \\
\lstick{\varphi \; \ket{0}_{s} } & \qw & \gate{\ket{\sum_{i=1}^{n}\left(\textbf{w}_i^x,\textbf{k}_i\right)}_s} & \meter
}
}
\caption{The quantum circuit for the whole life insurance with distribution loading $\mathcal{A}$ on the upper wires and the weighted sum $\mathcal{B}$ from equation (\ref{eq:8}) in the last register.}
\label{fig:whole_life_insurance}
\end{figure}
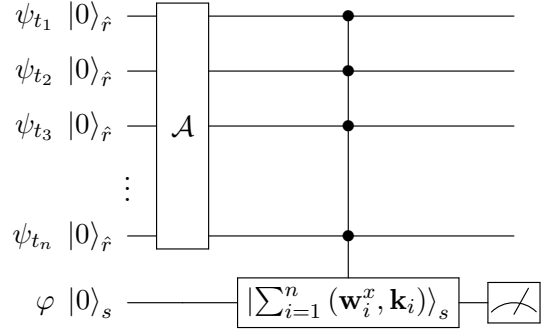

According to section \ref{sec:whole_life_intr} a whole life insurance payoff is a weighted sum of discount factors, where the weights are deterministic. A quantum implementation of a general weighted sum operator is demonstrated in \cite{stama20}, where it is used to price Asian options. Given the constant weights $w_0,\dots,w_{r-1}$ and a $r$-qubit register $\ket{\psi}_r$ in a basis state $\ket{k}_r = \ket{k_0\cdots k_{r-1}}$ there is an operator $\mathcal{B}$ which maps:
\begin{equation}
\label{eq:6}
\mathcal{B}\ket{k}_r\ket{0}_s = \ket{k}_r\ket{\sum_{j=0}^{r-1}w_jk_j}_s.
\end{equation}
This is a weighted sum of 1-bit integers $k_0,\dots,k_{r-1}$. If we want to implement the payoff from equation (\ref{eq:4}) the sum of $\hat{r}$-bit integers is needed. The solution is also given in \cite{stama20}: The weights $w_{t_i}^x$ have to be adjusted to fit the digits:
\[ \textbf{w}_{t_i}^x := (2^0\cdot w_{t_i}^x,\dots,2^{\hat{r}-1}\cdot w_{t_i}^x), i=1,\dots,n. \]
Hence all components of the payoff circuit are prepared. Let the marginal distributions of the discount factors $Z_{t_i}$ be represented by the $i^{th}$ part of the $r$-qubit register $\ket{\psi}_r$:
\begin{equation}
\label{eq:7}
\ket{\psi}_r = \ket{k_{t_1}}_{\hat{r}}\otimes\dots\otimes\ket{k_{t_n}}_{\hat{r}} = \ket{k_{{t_1},0}\dots k_{{t_1},\hat{r}-1}\dots k_{{t_n},0}\dots k_{{t_1},\hat{r}-1}},
\end{equation}
where $k_{t_i}$ are $\hat{r}$-bit integers. With binary vector $\textbf{k}_{t_i} = (k_{{t_i},0},\dots,k_{{t_i},\hat{r}-1})$ we finally get
\begin{equation}
\label{eq:8}
\begin{aligned}
\mathcal{B}(\mathcal{A}\otimes\mathcal{I}_{s})\ket{0}_{r+s} 
& = \ket{k}_r \ket{\sum_{i=1}^{n}\left(\textbf{w}_{t_i}^x,\textbf{k}_{t_i}\right)}_s, \\
\end{aligned}
\end{equation}
where the choice of $s$ must ensure the representability of the weighted sum. The quantum circuit is illustrated in figure \ref{fig:whole_life_insurance}.

\subsection{Dynamic Lapse}
\begin{figure}[ht]
\centerline{
\Qcircuit @C=1em @R=1em {
\lstick{\psi_{t_1} \; \ket{0}_{\hat{r}}} & \multigate{4}{\mathcal{A}} & \sgate{P_{t_1}}{5} & \qw & \qw & \qw & \qw & \ctrl{5} & \qw & \qw & \qw & \qw & \qw\\
\lstick{\psi_{t_2} \; \ket{0}_{\hat{r}}} & \ghost{\mathcal{A}} & \qw & \gate{P_{t_2}} & \qw & \qw & \qw & \qw & \ctrl{5} & \qw & \qw & \qw & \qw\\
\lstick{\psi_{t_3} \; \ket{0}_{\hat{r}}} & \ghost{\mathcal{A}} & \qw & \qw & \gate{P_{t_3}} & \qw & \qw & \qw & \qw & \ctrl{5} & \qw & \qw & \qw\\
\vdots & \nghost{\mathcal{A}} & & & & \ddots & & & & & \ddots \\
\lstick{\psi_n \; \ket{0}_{r_{\hat{r}}}} & \ghost{\mathcal{A}} & \qw & \qw & \qw & \qw & \qw & \qw & \qw & \qw & \qw & \ctrl{5} & \qw \\
\lstick{\varphi_{t_1} \; \ket{0} \hspace{2ex}} & \qw  &  \gate{P_{t_1}}  & \ctrlo{1} \qwx[-4] & \ctrlo{1} \qwx[-3] & \qw & \ctrlo{1} & \ctrl{5}  & \qw & \qw & \qw & \qw & \qw\\
\lstick{\varphi_{t_2} \; \ket{0} \hspace{2ex}} & \qw & \qw & \gate{P_{t_2}} & \ctrlo{1} & \qw & \ctrlo{1} & \qw  & \ctrl{4} & \qw & \qw & \qw & \qw\\
\lstick{\varphi_{t_3} \; \ket{0} \hspace{2ex}} & \qw & \qw & \qw & \gate{P_{t_3}} & \qw & \ctrlo{2} & \qw  & \qw & \ctrl{3} & \qw & \qw & \qw\\
\vdots & & & & & \ddots & & & & & \ddots\\
\lstick{\varphi_{t_n} \; \ket{0} \hspace{2ex}} & \qw & \qw & \qw & \qw & \qw & \gate{X} & \qw & \qw & \qw & \qw & \ctrl{1} & \qw \\
\lstick{\upsilon \; \ket{0}_s \hspace{1ex}} & \qw & \qw & \qw & \qw & \qw & \qw & \gate{X} & \gate{X} & \gate{X} & \qw & \gate{X} & \meter \\
& & & & & & & & & & & &  \\
\\
& 1 & 2.1 & \dots & & & 2.\textrm{n} & 3.1 & \dots & & & \textrm{3.n} & 4 
\gategroup{1}{2}{12}{2}{.8em}{_\}} 
\gategroup{1}{3}{12}{7}{.8em}{_\}}
\gategroup{1}{8}{12}{12}{.8em}{_\}} 
\gategroup{1}{13}{12}{13}{.8em}{_\}}\\
}
}
\caption{The quantum circuit for the dynamic lapse payoff with (1) Distribution loading (2.i) Multi-controlled linear amplitude function to model the lapse dynamics (3.i) Transmission of marginal distributions according to lapse events and (4) Measuring the expected value. Note that the open bullets indicate a control-on-zero while the filled bullets represent usual controls.}
\label{fig:dynamic_lapse}
\end{figure}
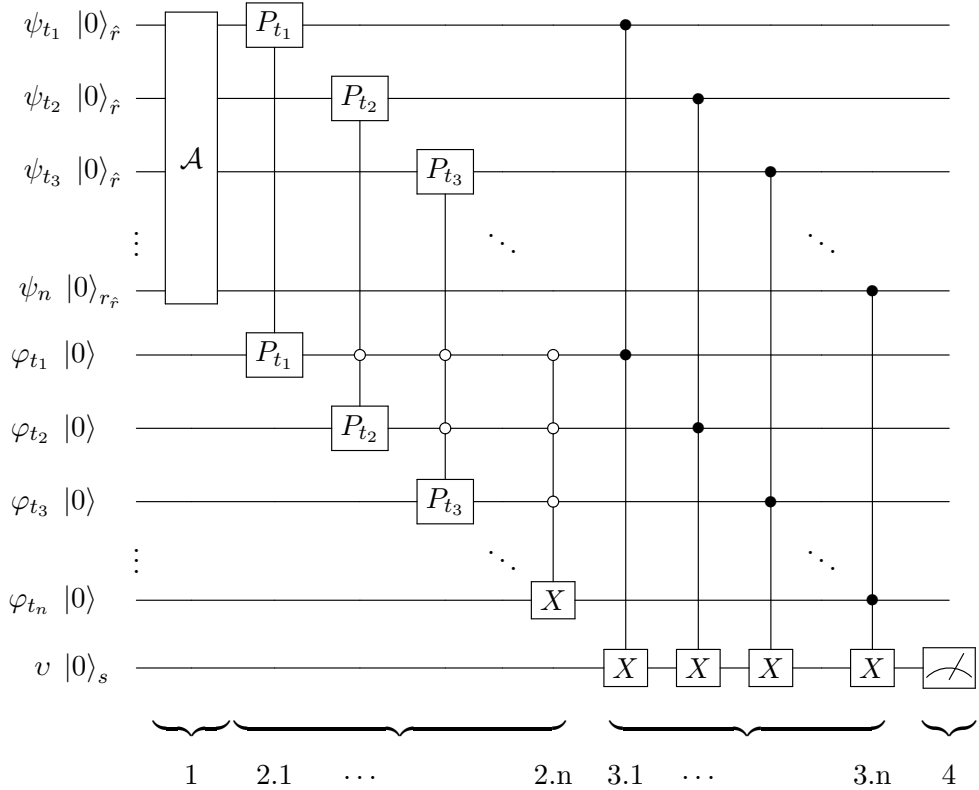

The components needed to implement the dynamic lapse payoff with specifications (\ref{eq:5}) are linear amplitude functions \cite{woerner19} for the Bernoulli-behavior with stochastic parameter (marginal lapse events) and multi-controlled versions of linear amplitude functions for the path-dependency of the stopping time (the process can only stop once). Additionally we use a sequence of Toffoli gates to sum up the discounted payments from the different time steps. Toffoli gates are also known as \textsc{ccnot} gates. Figure \ref{fig:dynamic_lapse} illustrates the corresponding quantum circuit.\\ 

We start with loading the distribution of the discount factors $(Z_{t_i}, i=1,\dots,n)$ to the register $\ket{\psi}_r$ partitioned like in (\ref{eq:7}). Later in the circuit we will use the first basis state $\ket{0}_{\hat{r}}$ of each time step $t_i$ for the share of contracts which are not canceled until $t_i$  (i.e. for $\PP({\tau > t_i})$), so we exclude it from the initial marginal distributions and the summations start with $k=1$:
\begin{equation}
\label{eq:9}
\ket{\psi_{t_i}}_{\hat{r}}: \ket{0}_{\hat{r}} \rightarrow \sum_{k=1}^{2^{\hat{r}}-1}\sqrt{a_k}\ket{k}_{\hat{r}}. 
\end{equation}

To ease notation, we assume identical probabilities $a_k$ for all $t_i$, but this can easily be implemented differently. Now we encode the stopping time. Therefore we add an $n$-qubit register $\ket{\varphi}_n$ to the circuit, i.e. one qubit for each time step. This register will finally be in a superposition of basis states $\{\ket{2^0}_n,\dots \ket{2^{n-1}}_n\}$, meaning that only $n$ out of $2^n$ available basis states are possible. The possible basis states are characterized by exactly one qubit being bin state $\ket{1}$. The position of this qubit indicates the time of contract cancellation. As described in section \ref{sec:distr_loading}, we follow the convention that the first digit is the least significant one.\\

The first time step can be implemented by a linear amplitude function $P_{t_1}$:
\[ P_1\ket{k_{t_1}}_{\hat{r}}\ket{0} = \ket{k_{t_1}}_{\hat{r}} \left( \sqrt{1-p_{t_1}(k_{t_1})}\ket{0} + \sqrt{p_{t_1}(k_{t_1})}\ket{1} \right). \]  
The last qubit $\ket{\varphi_{t_1}}$ in state $\ket{1}$ is interpreted as lapse event, i.e. the probability of cancellation at $t_1$ is $p_{t_1}(k_{t_1})$. Note that the lapse probability function $p_{t_i}$ must be defined on the domain $\{1,\dots,2^{\hat{r}}-1\}$ instead of $\RR$. Therefore we are implicitly assuming the mapping defined in (\ref{eq:aff_trans_kx}) with $[z_{min},z_{max}]$ being the ranges of $Z_{t_i}$.
For the following time steps, multi-controlled linear amplitude functions are needed because a lapse event can only occur if it hasn't already occurred:
\begin{equation}
\overline{\textsc{c}}^{i-1}P_{t_i}\ket{k_{t_i}}_{\hat{r}}\ket{l}_{i-1}\ket{0} = 
\begin{cases}
\ket{k_{t_i}}_{\hat{r}}\ket{l}_{i-1} \left( \sqrt{1-p_{t_i}(k_{t_i})}\ket{0} + \sqrt{p_{t_i}(k_{t_i})}\ket{1} \right), & \ket{l}_{i-1}=\ket{0}_{i-1} \\
\ket{k_{t_i}}_{\hat{r}}\ket{l}_{i-1}\ket{0}, & \textrm{else}.
\end{cases}
\end{equation}
At time step $t_i$, the $i-1$ precedent qubits $\ket{\varphi}_{i-1}$ serve as control qubits. In the given interpretation ($\ket{1}$ corresponds to lapse event) the control has to be negated, i.e. $P_{t_i}$ is applied if and only if all control qubits are in state $\ket{0}$ ("control-on-zero"). This can technically be achieved by applying \textsc{not} gates to all controls in front of the controlled function and reverting this afterwards. We are using $\overline{\textsc{c}}^i$ in our notation for a $i$-times control-on-zero. \\

Starting with $\ket{\varphi}_n=\ket{0}_n$, then applying $P_{t_1}$ in the first step and $\overline{\textsc{c}}^{i-1}P_{t_i}$ with $i=2,\dots,n-1$ in the following $n-2$ steps, the single qubits $\ket{\varphi_{t_i}}$ are successively entangled:
\begin{equation}
\begin{aligned}
\ket{\varphi_{t_1}} = & \sqrt{1-p_{t_1}(k_{t_1})}\ket{0} + \sqrt{p_{t_1}(k_{t_1})}\ket{1} \\
\ket{\varphi_{t_1}\varphi_{t_2}} = & \sqrt{p_{t_1}(k_{t_1})}\ket{10} + \sqrt{1-p_{t_1}(k_{t_1})}\sqrt{p_{t_2}(k_{t_2})}\ket{01} + \sqrt{1-p_{t_1}(k_{t_1})}\sqrt{1-p_{t_2}(k_{t_2})}\ket{00} \\
\cdots & \\
\ket{\varphi_{t_1}\dots\varphi_{t_{n-1}}} = & \sqrt{p_{t_1}(k_{t_1})}\ket{2^0}_{n-1} + \sqrt{1-p_{t_1}(k_{t_1})}\sqrt{p_{t_2}(k_{t_2})}\ket{2^1}_{n-1} \\
& + \sqrt{1-p_{t_1}(k_{t_1})}\sqrt{1-p_{t_2}(k_{t_2})}\sqrt{p_{t_3}(k_{t_3})}\ket{2^2}_{n-1} + \dots\\
& + \sqrt{1-p_{t_1}(k_{t_1})}\dots\sqrt{1-p_{t_{n-2}}(k_{t_{n-2}})}\sqrt{p_{t_{n-1}}(k_{t_{n-1}})}\ket{2^{n-2}}_{n-1} \\
& + \sqrt{1-p_{t_1}(k_{t_1})}\dots\sqrt{1-p_{t_{n-1}}(k_{t_{n-1}})}\ket{0}_{n-1}
\end{aligned}
\end{equation}
The last step in this entangling sequence is a multi-controlled-on-zero \textsc{not} gate, which eventually appends $\ket{1}$ to the register $\ket{\varphi}_{n-1}$ if none of the qubits in $\ket{\varphi}_{n-1}$ is in state $\ket{1}$:

\begin{equation*}
\overline{\textsc{c}}^{n-1}\textsc{not}\ket{l}_{n-1}\ket{0} = 
\begin{cases}
\ket{2^{n-1}}_{n}, & \ket{l}_{n-1}=\ket{0}_{n-1} \\
\ket{l}_{n-1}\ket{0}, & \textrm{else}.
\end{cases}
\end{equation*}

Hence we finally have the lapse distribution encoded in a $n$-qubit register:
\begin{equation}
\label{eq:final_lapse_reg_state}
\ket{\varphi}_n = \ket{\varphi_{t_1}\dots\varphi_{t_n}} = \sum_{i=1}^n \sqrt{ \PP(\{\tau=t_i\})} \ket{2^{i-1}}.
\end{equation}
I.e. $\ket{\varphi}_n$ is in the desired state as a superposition of only $n$ out of $2^n$ basis states, with the position of the single one indicating the time of contract cancellation. \\

In the final step of the payoff modeling we must calculate the sum of the marginal distributions according to the stopping time. We start with an additional $\hat{r}$-qubit register $\ket{\upsilon}_{\hat{r}}=\ket{0}_{\hat{r}}$. Then we are using the fact, that all basis states $\ket{l}_n$ of $\ket{\varphi}_n$ with probability greater than zero have exactly one qubit in state $\ket{1}$. Hence we don't need a real sum, but rather a "select case": If $\ket{l_{t_1}}=\ket{1}$ then transmit $\ket{\psi_{t_1}}_{\hat{r}}$ to $\ket{\upsilon}_{\hat{r}}$, else if $\ket{l_{t_2}}=\ket{1}$ then transmit $\ket{\psi_{t_2}}_{\hat{r}}$ and so on. This transmission is done via a sequence of $n\times\hat{r}$ Toffoli gates. For every time step $t_1,\dots,t_n$ there are $\hat{r}$ Toffolis with one control qubit bound to the stopping qubit $\ket{\varphi_{t_i}}$ and one each to the distribution qubits $\ket{\psi_{t_i,j}}, j=0,\dots,\hat{r}-1$. \\

We write $\textsc{ccnot}_i$ for a \textsc{ccnot} gate with second control qubit bound to the $i^{th}$ qubit of the second register:
\[ \textsc{ccnot}_i\ket{\psi_{t_i,j}}\ket{\varphi}_n\ket{\upsilon_j} = \ket{\psi_{t_i,j}}\ket{\varphi}_n\ket{\upsilon_j\oplus(\psi_{t_i,j}\wedge\varphi_{t_i})}, \]
where $\oplus$ denotes a \textsc{xor} and $\wedge$ is a logical \textsc{and}.\\

To see how the target register $\ket{\upsilon}_s$ evolves, we successively split the stopping time register into two parts with the first one including the basis state with the relevant control qubit is equal to one and the second one contains the rest. This split is only used for the formulas and has no impact on the circuit itself.      
\[ \ket{\varphi}_n = \sqrt{ p^{\tau}_{t_k}}\ket{2^{k-1}} + \sum_{i\neq k} \sqrt{ p^{\tau}_{t_i}}\ket{2^{i-1}}, \]
where $p^{\tau}_{t_i}:=\PP(\{\tau={t_i}\})$. When applying the first set of \textsc{ccnot}s, the target qubits are in state $\ket{0}$ and we get for $j=0,\dots,\hat{r}-1$:
\begin{equation}
\label{eq:11}
\textsc{ccnot}_1\ket{\psi_{{t_1},j}}\ket{\varphi}_n\ket{0} = \ket{\psi_{{t_1},j}}\sqrt{ p^{\tau}_1}\ket{2^0}\ket{\psi_{{t_1},j}} + \ket{\psi_{{t_1},j}} \sum_{i=2}^n \sqrt{ p^{\tau}_{t_i}}\ket{2^{i-1}} \ket{0}.
\end{equation}
After the first step, the target register $\ket{\upsilon}_{\hat{r}}$ is in state $\ket{0}_{\hat{r}}$ if and only if the contract is not canceled at the first time step. The "and only if" part follows from equation (\ref{eq:9}) because we didn't use the basis state $\ket{0}_{\hat{r}}$ for the distribution, i.e. $\ket{\psi_{t_i}}_{\hat{r}}\neq\ket{0}_{\hat{r}}$. For the next set of \textsc{ccnot}s we split the second part of (\ref{eq:11}) and so on. Finally the target register represents the distribution of the random variable defined in equation (\ref{eq:1}) with stopping time $\tau$ defined in section \ref{sec:dyn_lapse}:
\[ \ket{\upsilon}_s = \sum_{i=1}^n \sqrt{ p^{\tau}_{t_i}}\ket{\psi_{t_i}}_{\hat{r}}. \]

The size $s$ of the target register $\ket{\upsilon}_s$ must be chosen such that the resolutions of all marginal distributions can be represented. In our case with equal resolutions for all $t_i$, the target register simply needs the same resolution. Otherwise $s=\max r_{t_i}$ would be required. Given the payoff distribution encoded in the target register, we can eventually calculate the expected value as described in section \ref{sec:ev_calc}. We can also calculate other properties of the distribution, e.g. risk measures like VaR or CVaR \cite{woerner19}.

\section{Quantum Hardware Results}
\label{sec:hardware_results}
\begin{figure}[htbp]
\centering
\includegraphics[scale=1]{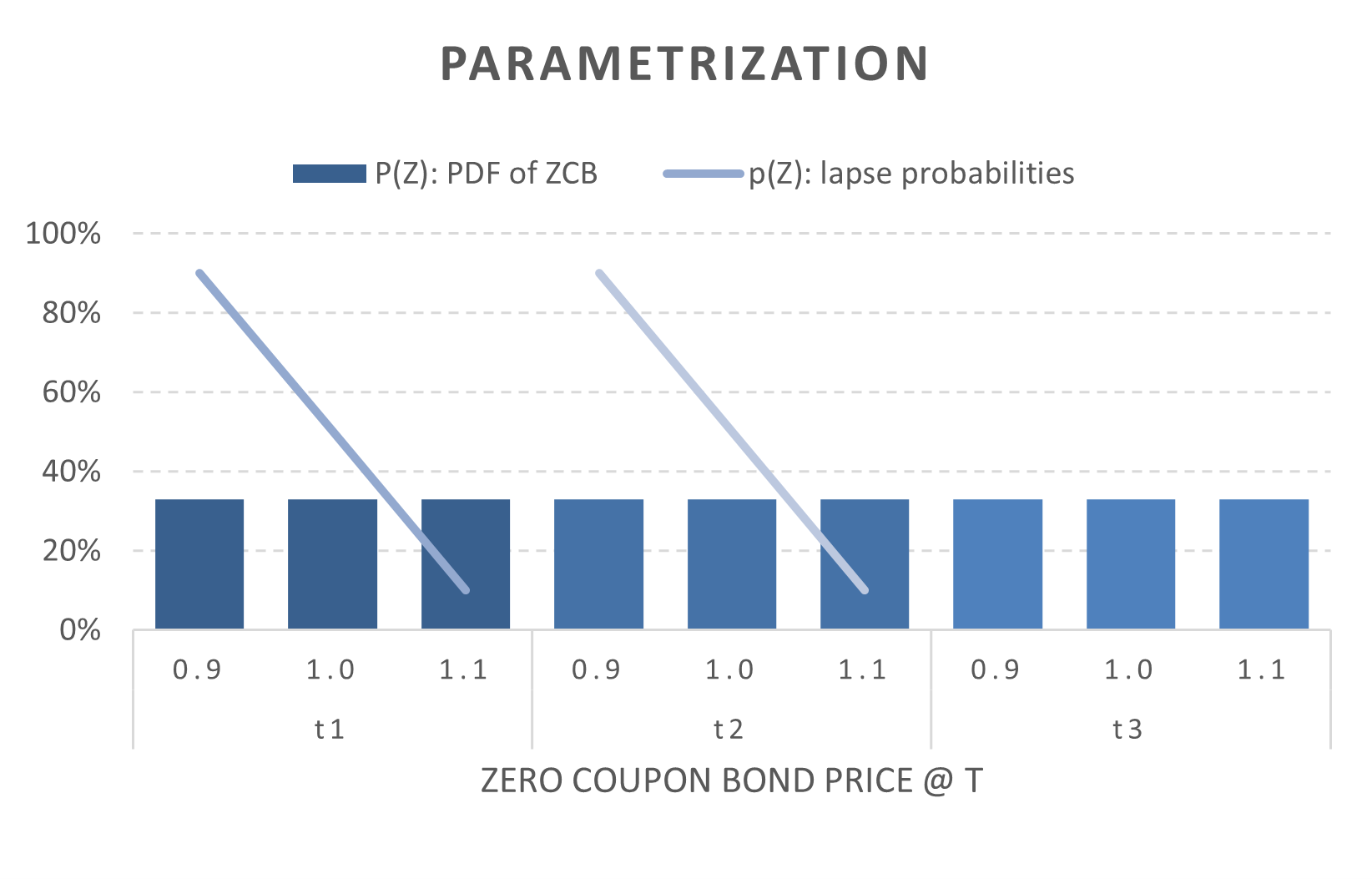}
\caption{Parametrization of the dynamic lapse circuit. Three time steps with two qubits each are provided for the probability distribution. The corresponding random variables $Z_i$ are assumed independent and uniformly distributed, considering that the basis states $\ket{00}$ are not used for the distribution. The dynamic lapse functions $p$ are linear and reflect that cancellation by customers is more likely in case of higher interest rates.} 
\label{fig:dyn_lapse_param}
\end{figure}
We focus on investigating the dynamic lapse circuit described in the previous section, illustrated in figure \ref{fig:dynamic_lapse}. The used payoff parametrization is shown in figure \ref{fig:dyn_lapse_param}: The probability distribution registers $\ket{\psi_{t_i}}_{\hat{r}}$, consisting of two qubits each, are initialized with uniform distributions. We proceed from the simplifying assumption that the zero coupon bond prices $Z_i$ are independent and identically distributed (iid), which in particular simplifies the distribution loading and makes the results easier to understand. Note that according to equation (\ref{eq:9}) only three out of four available basis states are used and hence the probabilities are $\nicefrac{1}{3}$ for each state. The "used" basis states correspond to the zero coupon prices $\{0.9, 1.0, 1.1\}$. The lapse probabilities $p$ are a linear function of the zero coupon bond prices with $p(0.9)=90\%, p(1.0)=50\%$ and $p(1.1)=10\%$, i.e. the higher the interest rate the higher the lapse probability. We don't use time dependent lapse behavior and hence omit the subscript of $p$. As the contract terminates in $t_i=3$, lapse probabilities are not needed here. Considering three time steps include all important cases for the desired lapse modeling: Unconditional lapse on the full portfolio at $t_i=1$, conditional and still interest rate dependent lapse for the remaining contracts at step two and eventually the incorporation of the survivors.\\

As a proof of concept we start this section by running the circuit on a simulator. Parts of the circuit are executed on real quantum hardware in the second part. 

\subsection{Simulator}
The following results are generated by a simulator, i.e. the figures have been calculated on a classical computer using Qiskit's quantum circuit simulator backend ("Aer Simulator") in its default configuration \cite{qiskit}.\\ 

\begin{figure}[htbp]
\centering
\includegraphics[scale=1]{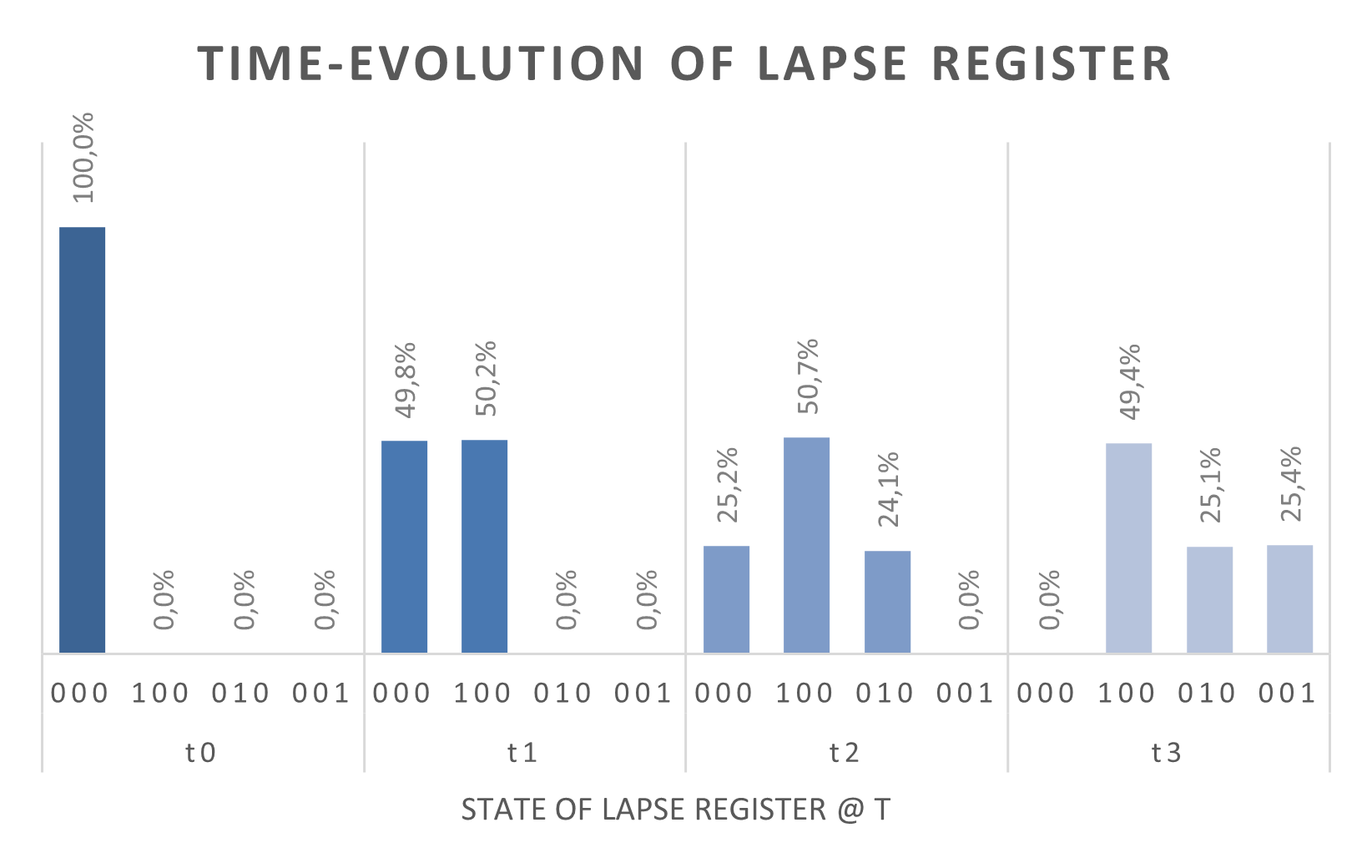}
\caption{Time-evolution of the lapse register $\ket{\varphi}_n$. The first basis state $\ket{000}$ represents the amount of active contracts and the time-evolution shows the successive transfer to the respective lapse dates $t_i=1,\dots,3$: At each time step $50\%$ of the remaining contracts are canceled. The interest rate dependency is not visible in these states.}
\label{fig:dyn_lapse_lapse_reg}
\end{figure}
\begin{figure}[htbp]
\centering
\includegraphics[scale=1]{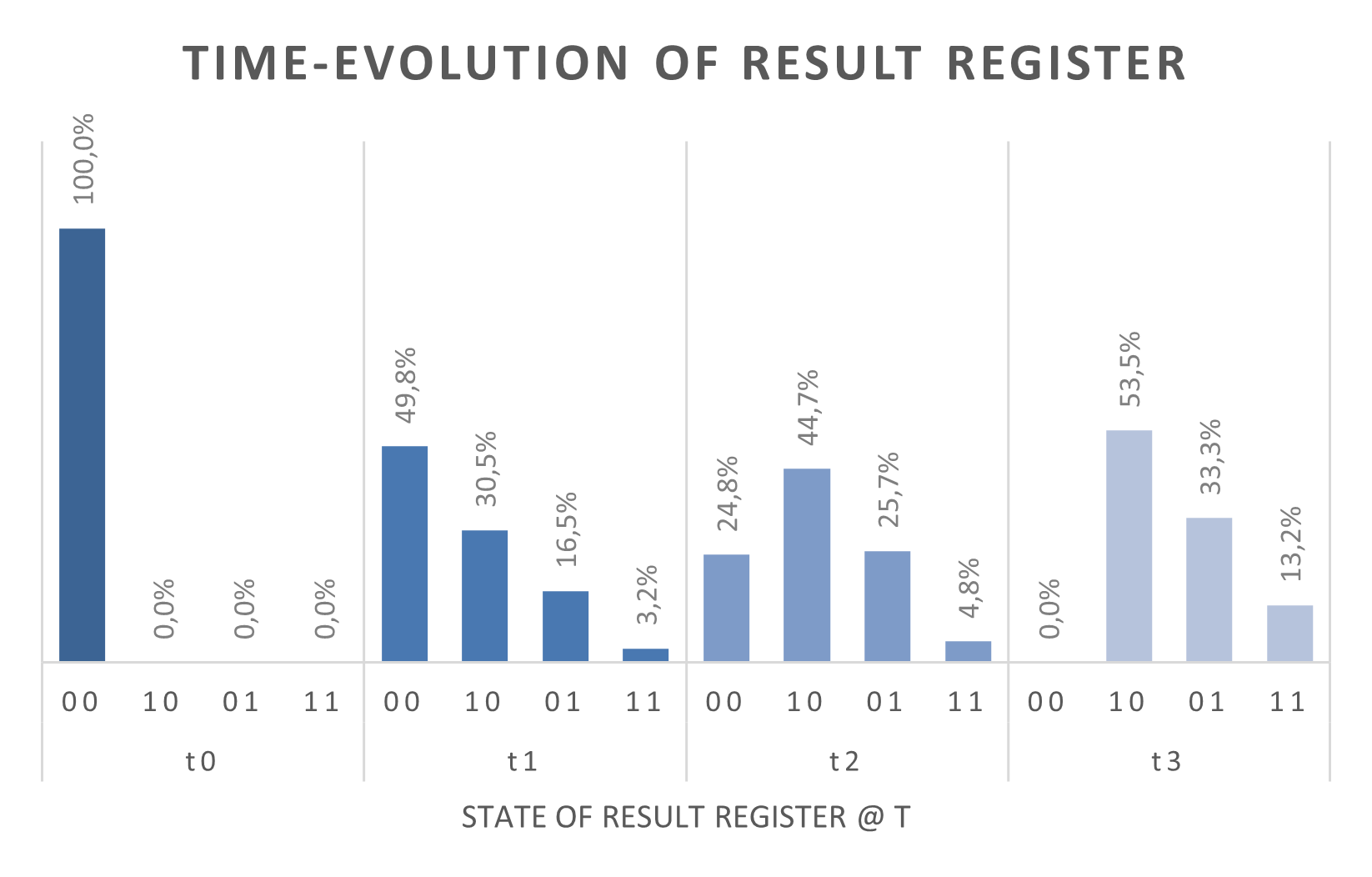}
\caption{Time-evolution of the result register $\ket{\upsilon}_s$. Like in figure \ref{fig:dyn_lapse_lapse_reg} the first basis state $\ket{00}$ represents the share of active contracts. The remaining states correspond to the zero coupon bond prices showing that cancellation is more likely in case of lower prices.}
\label{fig:dyn_lapse_result_reg}
\end{figure}

In figure \ref{fig:dyn_lapse_result_reg} we show measurement results of the lapse register $\ket{\varphi}_n$. The measurement has been performed at different steps of the circuit, which can be interpreted as different times $t_i=0,...,3$ of the stochastic process. Using the notation from the circuit illustration (figure \ref{fig:dyn_lapse_param}), the state of the lapse register has been measured before 2.1 (i.e. after initialization, we say $t_i=0$), before 2.2 ($t_i=1$) and so on. We see that the share of active contracts (state $\ket{000}$) decreases by $50\%$ in each time step until the remaining part is shifted to $\ket{001}$ at $t_i=3$. Note that eventually there are only three possible basis states, i.e. a contract can only be canceled once like formally shown in equation (\ref{eq:final_lapse_reg_state}).\\ 

Figure \ref{fig:dyn_lapse_result_reg} shows the evolution of the result register $\ket{\upsilon}_s$, again measured at different steps: Before 3.1 ($t_i=0$), before 3.2 ($t_i=1$) and finally at step 4. The state $\ket{00}$ represents the share of active contracts and hence corresponds to the state $\ket{000}$ of the lapse register. The other states show the distribution of the canceled contracts to the different zero coupon bond prices. If for example the zero coupon price at $t_i=1$ is $0.9$, the corresponding lapse probability is $p(0.9)=90\%$ and hence $90\%$ of the contracts are canceled. As the probability for a zero coupon bond price $Z_1=0.9$ amounts to $\nicefrac{1}{3}$, the share of contracts which are canceled in $t_i=1$ in case of $Z_1=0.9$ amounts to $\PP(Z_1=0.9)\cdot p(0.9)\approx 30\%$. After the first cancellation process there are $50\%$ of active contracts left. As the bond price distribution does not change (due to iid property), the cancellation works equivalently to first time step, but only for the remaining contracts. Hence the probabilities for the basis states $k=1,\dots,3$ are each increased by $50\%\cdot\PP(Z_2=k)\cdot p(k)$. In the last step, the contracts terminate as scheduled and hence there is no interest rate dependency. The remaining $25\%$ of active contracts (which is equal to the probability of "surviving" time steps $1$ and $2$) are uniformly distributed to the bond price basis states, i.e. each probability is increased by ca. $8\%$. Eventually the customer behavior lowers the final payoff in this basic example: While the expected value would be $1$ for a uniform distribution, the interest rate dependent lapse shifts weight to lower bond prices leading to the distribution shown in figure \ref{fig:dyn_lapse_result_reg}. For the given parametrization the expected value amounts to ca. $0.96$ after dynamic lapse. 

\subsection{Real Hardware}
The simulator results from the previous section show that the exemplary dynamic lapse circuit would deliver the expected payoff distribution on a sufficiently large fault-tolerant quantum computer. In this section we show the results from running the circuit on the \textit{IBM Q System One} in Ehningen. Its processor type is \textit{Falcon r5.11} with 27 qubits. A quantum processor is also called QPU (quantum processing unit).\\

Before we start, we briefly introduce the concept of \textit{universality}. For details see \cite{kaye07}, sections 4.3 and 4.4. A set of quantum gates is called \textit{universal} if any operator can be "replicated" by a quantum circuit assembled from this set. Replicated in this context means that the original gate can be approximated with arbitrary accuracy. One can show that (1) universal sets exist and (2) the replication of any gate can be done efficiently if the universal set satisfies certain conditions. In practice, every QPU has a universal set of gates which it can actually execute. These gates are called \textit{basis gates} of the processor. Hence, to run a circuit on a certain QPU, it needs to be \textit{transpiled} to the basis gates, i.e. each gate of the original circuit must be replicated by the basis gates. The basis gates of the Falcon QPU are: \textsc{cnot}, Identity, $R_z(\theta)$, $\sqrt{\textsc{not}}$, \textsc{not}. Hence the set consists of one 2-qubit gate and four 1-qubit gates. The transpiling is usually done by a compiler and in high level programming, the coder usually does not care about the details. However, given the early stage of quantum computing, this step may influence the results and is hence itself a research topic. For this work, we did not concentrate on compiling and simply used the Qiskit standards.  \\ 

When we investigate the suitability of a quantum circuit to be run on real hardware, there are basically two crucial properties: The \textit{width} and the \textit{depth} of the circuit. The circuit's width determines how many qubits are needed ("vertical size") and the depth is defined as the longest path through the circuit ("horizontal size"). As QPUs allow for real parallelism, the depth presumes a fully parallelized circuit.\\ 

Given a concrete quantum processor, the number of qubits must be greater or equal to the width of a circuit. Otherwise the circuit cannot be run at all so that this can be regarded as the necessary condition for executing a circuit on a certain QPU. This property is also referred to as the QPU's \textit{scale}. Whether the processor is capable of the circuit's depth depends on the \textit{quality} of the qubits. There are different ways for measuring the quality, but the attributes \textit{coherence time} and \textit{gate fidelity} always play an important role. Coherence time tells us how long a qubit retains its information. A perfectly isolated system maintains coherence indefinitely but since the heart of quantum computing is the manipulation of qubits, they must interact with the environment which in turn  causes decoherence. If decoherence progresses, the state of the system can move away from the encoded state and hence get less and less meaningful. The gate fidelity tells us something about the precision of a gate. The higher the gate fidelity, the higher the confidence that a gate prepares the theoretically expected state. Obviously the deeper the circuit the higher the requirements on the QPU quality, i.e. on coherence time and gate fidelity. If the QPU quality is not sufficient for a given circuit, the circuit can still be executed but the result will be meaningless. Hence we can call the QPU quality the sufficient condition for running a circuit.\\

Wack et al \cite{wack21} classify the scale, the quality and the speed to be the key attributes for measuring performance of near-term quantum computers. In addition to the scale and the quality mentioned above, the \textit{speed} is defined as the time needed for executing a circuit. Of course the speed is also important for getting a quantum advantage.\\

Now we will examine the necessary and the sufficient condition for executing the exemplary dynamic lapse circuit described in the previous section on the Falcon r5.11 processor. 

\subsubsection{Necessary Condition}
To assess the necessary condition we need to compare the circuits width with the QPU's scale. The exemplary dynamic lapse circuit requires 17 qubits: 3x2 qubits for the underlying distribution, 3 for the stopping time, 2 for the payoff distribution and 6 ancillas for the linear amplitude functions. Hence the necessary condition is fulfilled and it can be executed on the Falcon processor with 27 qubits. Note that this is the most simplified but still meaningful case. Also note that the actual width of the circuit is 19, because we additionally need two classical bits for storing the measurement outcome of the result register.

\subsubsection{Sufficient Condition}
To assess the sufficient condition we need to compare the circuits depth with the QPU's quality. The depth of a circuit can be considered on different levels. For example, counting the operations in figure \ref{fig:dynamic_lapse} leads to a high level depth of 10: 1 gate for distribution loading ($\mathcal{A}$), 3 gates for the stopping time ($P_{t_i}$) and 3x2 gates for the  the sum of time steps (\textsc{ccnot}). Obviously this figure is not very useful, because the high level gates have varying \textit{costs}. The costs of a gate are a measure for the quality requirements. To get a comparable figure we first transpile the circuit to the basis gates of the processor and then assign a \textit{cost} to each basis gate. The latter is important, because even the basis gates have different costs. Lee et. al \cite{lee06} define the cost of a quantum gate as the number of basic physical operations needed for its implementation. Table \ref{tab:bg_cost} shows the costs they propose for Falcon's basis gates. Note that the 2-qubit gate \textsc{not} is much more expensive than the single qubit rotations.\\

To get an idea of how expensive the different operations of the dynamic lapse circuit are, we show the depth and the costs at each step in the table \ref{tab:dynlapse_cost}. The table also shows the number of the respective basis gates along the longest path always assuming maximum parallelization. The figures are cumulative, e.g. step 3.1 represents the entire circuit until and including this step. While the implementation of the uniform distribution is relatively cheap, we see immediately that the linear amplitude function in step 2.1 and particularly its controlled version in step 2.2 is very expensive. The summation in step 3 does not increase the costs a lot, because large parts can be done in parallel to step 2.\\

\begin{figure}
\centering
\includegraphics[scale=.7]{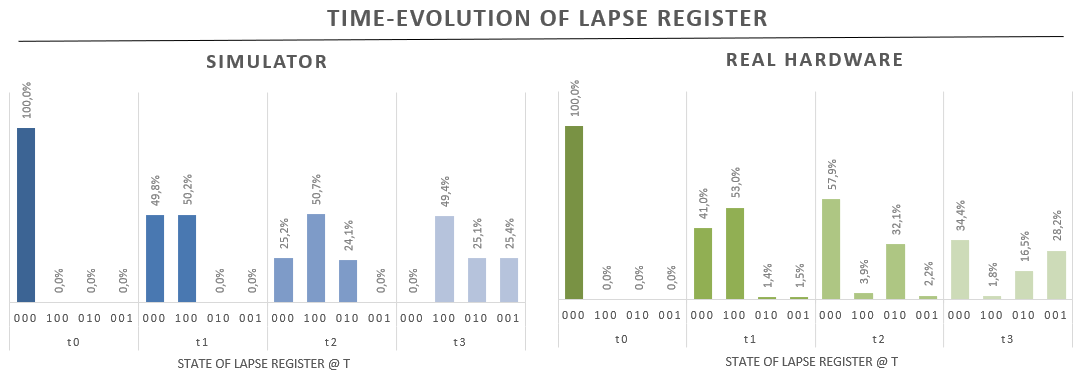}
\caption{Comparison of simulator and real hardware results for the lapse register. We can see that the first lapse event delivers results which are close to the theoretical expectation. After a tenfold increase of the costs by the controlled linear amplitude function at step 2.2, the QPU returns more or less meaningless results.}
\label{fig:dyn_lapse_lapse_reg_real_hardware}
\end{figure}

The results shown in figure \ref{fig:dyn_lapse_lapse_reg_real_hardware} lead to an assessment of the QPU quality. Measuring after step 2.1 delivers results which are close the simulator (see the state of lapse register at $t_1$). That shows that the QPU quality is at least sufficient for circuits with costs around 300. The sharply increased costs at step 2.2 leads to more or less meaningless results at step 2.2 (lapse register at $t_2$). Since the state of the qubit system is already meaningless at this step, we don't show the measurement results for step 3. Note that we did not concentrate on result improvements by circuit optimization. Potential next steps for improving the results could be the application of error mitigation techniques and searching for alternative implementations of the stopping time.   

\begin{table}[htbp]
\centering
\begin{tabular}{c|c}
\textbf{Basis gate} & \textbf{Cost} \\
\hline
\textsc{cnot} & 5 \\
Identity & 1 \\
$R_z(\theta)$ & 1 \\
$\sqrt{\textsc{not}}$ & 1 \\
\textsc{not} & 1 \\
\end{tabular}
\caption{The costs of Falcon's basis gates according to \cite{lee06}, where the cost of a quantum gate is defined as the number of basic physical operations needed for its implementation. Obviously the number of \textsc{cnot}s drives the total cost. The higher the cost the higher the requirements on the QPU's quality.}
\label{tab:bg_cost}
\end{table}

\begin{table}
\centering
\begin{tabular}{c||c|c|c|c||c|c}
\textbf{Step} & \textsc{cnot} & $R_z(\theta)$ & $\sqrt{\textsc{not}}$ & \textsc{not} & \textbf{Depth} & \textbf{Cost} \\
\hline
\textbf{1} & 1 & 4 & 4 & 0 & 9 & 13 \\ \hdashline
\textbf{2.1} & 47 & 47 & 26 & 2 & 122 & 310 \\
\textbf{2.2} & 485 & 467 & 133 & 1 & 1086 & 3026 \\
\textbf{2.3} & 488 & 471 & 134 & 1 & 1094 & 3046 \\ \hdashline
\textbf{3.1} & 490 & 470 & 133 & 1 & 1094 & 3054 \\
\textbf{3.2} & 494 & 475 & 134 & 1 & 1104 & 3080 \\
\textbf{3.3} & 498 & 480 & 135 & 1 & 1114 & 3106 \\
\end{tabular}
\caption{Cumulative counts and costs for the longest path until and including step $N$. We can see that the distribution loading is relatively cheap while the stopping time is the most expensive part.}
\label{tab:dynlapse_cost}
\end{table}

\section{An Overview of current Quantum Computing Technology}
\label{sec:current_tech}
We are in the era of Noisy Intermediate-Scale Quantum (NISQ) technology. The term was introduced by John Preskill in 2018 \cite{presk18} and should emphasize that currently available quantum computers are neither fault-tolerant nor do they have a large number of qubits. His main message is that "we may feel confident that
quantum technology will have a substantial impact on society in the decades ahead, but we cannot be nearly so confident about the commercial potential of quantum technology in the near term, say the next five to ten years." He concludes that fault-tolerance must be the essential longer-term goal and that the necessary efforts to reach this goal are worth it. The recently updated IBM development road map \cite{gambetta22} covers the period until 2025. During this period they predict an increasing number of qubits from currently less than 100 up to 4K+ accompanied by error mitigation and growing software development frameworks. Actual error correction and scaling to 10K-100K qubits is targeted "beyond 2026".

\subsection{Software}
Several quantum computing developments libraries are available. The probably most popular ones are IBM's Qiskit \cite{qiskit}, Google's Cirq \cite{cirq} and Microsoft's Quantum Development Kit \cite{qdk}. All of them are open-source. While Qiskit and Cirq are based in Python, Microsoft introduces a dedicated quantum programming language called Q\#. We won't analyze the details and differences of these frameworks but want to demonstrate the general idea of how they work. In particular we want to emphasize that everyone can start implementing quantum algorithms right away and thanks to cloud quantum computers, even the technical barriers to running it on real hardware are rather small. Although programming level is still low (meaning that we are operating on qubit level), the mentioned frameworks provide toolkits which allow for comfortable circuit assembling based on a growing library of generic circuits.\\

The following code example shows that Qiskit literally works like a kit: We gather the needed building blocks, assemble them as we like and finally execute the circuit on a provided backend. In this case we are using a simulator, i.e. the quantum computer is simulated on a classical one. To run the circuit on real hardware, we only need to get another backend, i.e. just one line of code needs to be adjusted.

\begin{lstlisting}
from qiskit import Aer, QuantumRegister, ClassicalRegister, QuantumCircuit, execute
from qiskit.circuit.library import WeightedAdder

# initialize adder
weighted_adder = WeightedAdder(num_state_qubits=2, weights=[1, 1])

# build empty quantum circuit with proper number of qubits
qreg = QuantumRegister(size=weighted_adder.num_qubits)
creg = ClassicalRegister(size=weighted_adder.num_sum_qubits)
circuit = QuantumCircuit(qreg, creg)

# assemble circuit from available gates/circuits
circuit.h(qreg[0])
circuit.h(qreg[1])
circuit.append(weighted_adder, qreg)

# add a measurement to the adder's sum qubits
circuit.measure(qreg[weighted_adder.num_state_qubits:
							   weighted_adder.num_state_qubits
							   +weighted_adder.num_sum_qubits], creg)

# get simulator and run the circuit 1000 times
simulator = Aer.get_backend('aer_simulator')
job = execute(circuit, simulator, shots=1000)
result = job.result()

# print the result statistics
print(result.get_counts(circuit))
# output: {'10': 502, '00': 231, '01': 267}
\end{lstlisting}

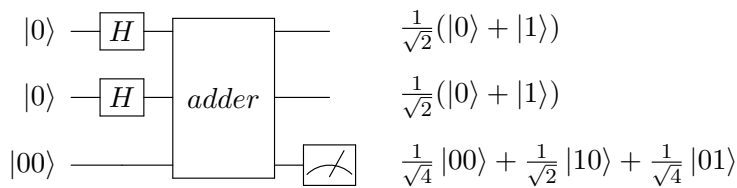
\begin{figure}[ht]
\centerline{
\Qcircuit @C=1em @R=1em {
\lstick{\ket{0}} & \gate{H} & \multigate{2}{adder} &\qw & \rstick{ \frac{1}{\sqrt{2}}(\ket{0}+\ket{1})} \\   
\lstick{\ket{0}} & \gate{H} & \ghost{adder} &\qw & \rstick{ \frac{1}{\sqrt{2}}(\ket{0}+\ket{1})} \\ 
\lstick{\ket{00}} & \qw & \ghost{adder}  & \meter & \rstick{\frac{1}{\sqrt{4}}\ket{00}+\frac{1}{\sqrt{2}}\ket{10}+\frac{1}{\sqrt{4}}\ket{01}} \\ 
}
}
\caption{Circuit illustration of the code fragment including the generic adder gate, which encapsulates a sequence of (controlled) \textsc{not}s.}
\label{fig:qiskit_xmp}
\end{figure}

For the sake of completeness: We are adding two qubits in state $H(\ket{0})=\frac{1}{\sqrt{2}}(\ket{0}+\ket{1})$ each weighted with $1$, where $H$ denotes the Hadamard gate. Hence the possible results are $0+0, 1+0, 0+1, 1+1$, each with probability \nicefrac{1}{4}. The circuit is illustrated in figure \ref{fig:qiskit_xmp}. Please note that the weighted adder is an example of available generic circuits. Under the hood, it is actually a sequence of \textsc{not}s, \textsc{cnot}s and \textsc{ccnot}s. The components of the dynamic lapse circuit from figure \ref{fig:dynamic_lapse} can also be encapsulated to contribute three generic circuits to the library: (1) The multivariate distribution, (2) the stopping time and (3) the stopped process. 

Finally note that although there are existing and growing frameworks the status quo of quantum software development is still very basic. For example there are no data structures like lists or dictionaries and there are only very few quantum algorithms available. 

\subsection{Hardware}
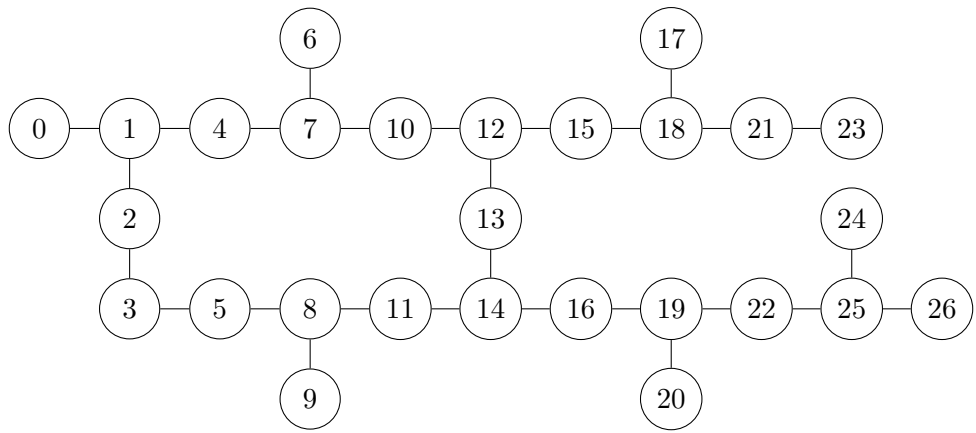
\begin{figure}[htbp]
\centering
\begin{tikzpicture}[node distance=1.2cm, minimum size=0.8cm, main/.style={draw, circle}] 
\node[main] (0) {0}; 
\node[main] (1) [right of=0] {1};
\node[main] (2) [below of=1] {2};
\node[main] (3) [below of=2] {3};
\node[main] (4) [right of=1] {4};
\node[main] (5) [right of=3] {5};
\node[main] (7) [right of=4] {7};
\node[main] (6) [above of=7] {6};
\node[main] (8) [right of=5] {8};
\node[main] (9) [below of=8] {9};
\node[main] (10) [right of=7] {10};
\node[main] (11) [right of=8] {11};
\node[main] (12) [right of=10] {12};
\node[main] (13) [below of=12] {13};
\node[main] (14) [below of=13] {14};
\node[main] (15) [right of=12] {15};
\node[main] (16) [right of=14] {16};
\node[main] (18) [right of=15] {18};
\node[main] (17) [above of=18] {17};
\node[main] (19) [right of=16] {19};
\node[main] (20) [below of=19] {20};
\node[main] (21) [right of=18] {21};
\node[main] (22) [right of=19] {22};
\node[main] (23) [right of=21] {23};
\node[main] (24) [below of=23] {24};
\node[main] (25) [right of=22] {25};
\node[main] (26) [right of=25] {26};

\draw (0) -- (1);
\draw (1) -- (4);
\draw (4) -- (7);
\draw (7) -- (10);
\draw (7) -- (6);
\draw (10) -- (12);
\draw (12) -- (13);
\draw (12) -- (15);
\draw (15) -- (18);
\draw (18) -- (17);
\draw (18) -- (21);
\draw (21) -- (23);
\draw (1) -- (2);
\draw (2) -- (3);
\draw (3) -- (5);
\draw (5) -- (8);
\draw (8) -- (9);
\draw (8) -- (11);
\draw (11) -- (14);
\draw (14) -- (13);
\draw (14) -- (16);
\draw (16) -- (19);
\draw (19) -- (20);
\draw (19) -- (22);
\draw (22) -- (25);
\draw (25) -- (24);
\draw (25) -- (26);
\end{tikzpicture}

\caption{Heavy-hexagon architecture of the IBM Q System One with 27 qubits, which was used for the experiments in section \ref{sec:hardware_results}. The processor type is \textit{Falcon r5.11}. According to IBM, this topology reduces error rates and should help reaching the ultimate goal of demonstrating fault-tolerance \cite{heavyhex}.}
\label{fig:heavy_hex_lattice}
\end{figure}

In our work we focus on \textit{gate-based} quantum computers. The underlying principle is the description of interactions between qubits and gates in a \textit{quantum circuit model}. This is closely related to classical computers with of course different kind of memory units (qubits vs bits) and different kind operating units (quantum vs. classical gates). Many of the well-known quantum algorithms like Shor's factoring algorithm \cite{shor94} or Grover's search algorithm \cite{grover96} are expressed in a circuit model. IBM's Q System One is a gate-based quantum computer with 27 qubits using heavy-hexagon architecture \cite{heavyhex}, illustrated in figure \ref{fig:heavy_hex_lattice}.\\ 

In contrast to gate-based devices, \textit{Quantum Annealing} requires a problem to be formulated as quadratic unconstrained binary optimization problem (QUBO). This is less generic than the gate-based approach, but currently available hardware is further advanced. For example, D-Wave Systems claims to provide the most powerful quantum computer in the world with 5K qubits. We won't dig deeper into the quantum annealing and refer to \cite{kadowaki98} for the theoretical foundation and to D-Wave's documentation \cite{dwave} for an application manual.\\

We only gave a very short overview of the current status of quantum technology. The main points we want to emphasize is that (1) both soft- and hardware are in their infancies (2) there is a lot of activity in both areas and (3) the best solutions are still to be found. It is not clear whether a universal gate-based computer is needed or if a quantum annealer better solves practical problems. Likewise we don't know if millions of qubits are necessary or if a small number of high quality would be more beneficial. Also hybrid approaches with QPUs, GPUs and CPUs (quantum, graphical and central processing units) are promising, where each unit is doing what it is best suited for.

\section{Conclusion}
Quantum computing is still in its infancy. The basis of quantum mechanics have been developed in the first half of the 20th century and the first approaches to quantum computing date back to the 80s. In the following decades several promising but from the lack of hardware, mostly theoretical results have been achieved, e.g. Shor's prime factoring algorithm. However in recent years activities have increased and some milestones have been reached. Amongst these the introduction of the first circuit-based commercial quantum computer in 2019, IBM's Q System One. In 2021 a unit of this system has been deployed in Germany, being the first quantum computer in Europe. IBM updated their ambitious quantum road map in May 2022 showing a combined path of soft- and hardware development until 2025 and beyond. Other big players like Google ("Google Quantum AI") or Microsoft increase their efforts and are already providing quantum software development kits. McKinsey \cite{mckinsey22} counts a total amount of $\$31$ billion of funding for quantum technologies since 2001 and expect the impact on financial industry "incremental" until 2030 but "significant" until 2035. It is common sense that the question is less \textit{whether}, but \textit{when} quantum computing becomes a game changer in many areas.\\

There are different aspects to be considered for the insurance industry. In this paper we focused on the quadratic speed-up of Monte Carlo simulations and how this could be applied to valuate life insurance contracts. We see a lot of work ahead of us, but some quantum circuits which model "insurance-relevant" behavior like general optionality or dynamic lapse are already available. Hence we consider the goal of being able to dynamically valuate an insurance balance sheet achievable. Of course this is subject to the invention of a fault-tolerant quantum computer. We think that soft- and hardware research should ideally progress jointly so that they mutually fulfill their needs. Aside from direct application for simulation or optimization, insurance companies will benefit from developing quantum computing know-how. As insurance services affect many areas of life, most future impacts of quantum computing will somehow also affect insurance business. This could be topics like cyber security, medical products or logistics. At the current status we finally conclude that a moderate investment in quantum computing is profitable for insurance companies. To be able to benefit from future inventions, building up know-how and tracking ongoing developments should start right now.

\end{document}